\documentclass[prb,twocolumn,10pt,aps]{revtex4-2}

\usepackage[utf8]{inputenc}
\usepackage[english]{babel}
\usepackage{indentfirst}

\usepackage[final]{showkeys}
\usepackage{graphicx}
\usepackage{subfig}
\usepackage{amssymb}
\usepackage{latexsym}
\usepackage{amsthm}
\usepackage{amsmath}
\usepackage{amsfonts}
\usepackage{mathtools}
\usepackage{pgf,tikz}
\usepackage{mathrsfs}
\usepackage{epstopdf}
\usetikzlibrary{arrows}
\usepackage{enumerate}
\usepackage{epigraph}
\usepackage{calrsfs}
\usepackage[normalem]{ulem}
\usepackage[a4paper, left=1.5cm, right=1.5cm, top=2cm, bottom=2cm]{geometry}
\usepackage{floatpag}
\DeclareUnicodeCharacter{2212}{-}
\DeclareUnicodeCharacter{03B4}{-}

\begin{document}

\title{Ensemble inequivalence in long-range quantum systems}

\author{Nicol\`o Defenu}
\email{ndefenu@phys.ethz.ch}
\affiliation{Institute for Theoretical Physics, ETH Z$\ddot{u}$rich, Wolfgang-Pauli-Str. 27, 8093 Z$\ddot{u}$rich, Switzerland}
\author{David Mukamel}
\affiliation{Department of Physics of Complex Systems, Weizmann Institute of Science, Rehovot
7610001, Israel}
\author{Stefano Ruffo}
\affiliation{SISSA and INFN Sezione di Trieste, Via Bonomea 265, I-34136 Trieste, Italy}
\affiliation{Istituto dei Sistemi Complessi, Consiglio Nazionale delle Ricerche, Via Madonna del Piano 10, I-50019 Sesto Fiorentino, Italy}

\begin{abstract}
Ensemble inequivalence, i.e. the possibility of observing different thermodynamic properties depending on the statistical ensemble which describes the system, is one of the hallmarks of long-range physics, which has been demonstrated in numerous classical systems.
Here, an example of ensemble inequivalence of a long-range \emph{quantum} ferromagnet is presented. While the $T=0$ microcanonical quantum phase-diagram  coincides with that of the canonical ensemble, the phase-diagrams of the two ensembles are different at finite temperature. This is in contrast with the common lore of statistical mechanics of systems with short-range interactions where thermodynamic properties are bound to coincide for macroscopic systems described by different ensembles. The consequences of these findings in the context of atomic, molecular and optical (AMO) setups are delineated.
\end{abstract}
\maketitle

\emph{Introduction:} In recent years the interest of the research community in the equilibrium and dynamical behaviour of long-range interacting quantum systems has experienced a unprecedented surge. Part of this enthusiasm stems from recent developments in the control, manipulation and observation of atomic, molecular and optical (AMO) systems, where long-range interactions within the microscopic components of the system are prevalent\,\cite{lukin2003trapping, saffman2010quantum, britton2012engineered, bloch2008many, blatt2012quantum, monroe2021programmable, mivehvar2021cavity,defenu2023longrange}. Conventionally, we refer to a many-body system as long-ranged if its two-body interaction potential $V(r)$  decays as a power law of the distance $r$, $V(r)\propto r^{-\alpha}$, with sufficiently small and positive $\alpha$.  

The system's phenomenology is heavily influenced by the exponent $\alpha$. For $\alpha>\alpha_{*}$, with $\alpha_{*}$ a universal threshold value, the critical behavior mirrors that of systems with short-range interactions, while for $d<\alpha<\alpha_{*}$ in $d$ dimensions the universal scaling near the phase transition is modified by the long-range couplings\,\cite{angelini2014relations,defenu2015fixed,defenu2016anisotropic,defenu2017criticality,defenu2020criticality}. In the strong long-range regime ($\alpha<d$), where traditional thermodynamics does not apply, rescaling the interaction strength by an appropriate size dependent factor, i.e. the Kac's rescaling, restores energy extensivity but leaves most thermodynamic functions non-additive. Notably, it has been demonstrated in numerous classical systems that this regime exhibits the appearence of \emph{quasi-stationary states} (QSSs)\,\cite{dauxois2002hamiltonian,campa2009statistical,campa2014physics,levin2014nonequilibrium} and \emph{ensemble inequivalence}\,\cite{barre2001inequivalence}, two of the hallmarks of long-range physics. QSSs are metastable configurations of the out-of-equilibrium dynamics, whose lifetimes diverge with the system size\,\cite{dauxois2002hamiltonian,campa2009statistical,campa2014physics}, while \emph{ensemble inequivalence}  results in differing properties across thermodynamic ensembles and is the focus of the present manuscript. 

The quantum statistical mechanics of strong long-range interacting systems is to a large extent unexplored with few notable exceptions, which have been identified following the classical physics chart.In fact, theoretical evidence on quantum QSSs\,\cite{kastner2011diverging,schutz2014prethermalization,schutz2016dissipation,mori2019prethermalization}, which recently found a unified explanation based on the quasi-particle 
spectrum in strong long-range systems\,\cite{defenu2021metastability}, should be compared with ergodicity breaking in classical systems, which is especially 
relevant in the microcanonical ensemble~\cite{Schreiber,Borgonovi}. Ensemble inequivalence has also been discussed in the context of the finite temperature transition of quantum mechanical models~\cite{kastner2010nonequivalence,kastner2010nonequivalence2,russomanno2021chaos}. With the rise of quantum simulators featuring native long-range interactions~\cite{defenu2023longrange}, it has become crucial to understand the impact of this significant phenomenon in the vicinity of a quantum critical point.

In this paper, a genuine example of quantum ensemble inequivalence is presented, where the canonical and microcanical phase diagrams are different. This is done by analyzing a quantum model with long-range fully connected interactions and multi-spin couplings which is known to exhibit a $T=0$ paramagnetic to ferromagnetic transition~\cite{Delre}. The transition line is composed of first-order and
second-order segments separated by a tricritical point. 
We show that, while the two ensembles yield the same $T=0$ phase-diagram, they result in different finite temperature phase-diagrams.

The relevance of this study is particularly evident nowadays, as the AMO community is pushing the investigation of quantum many-body systems toward the control of multi-body interactions\,\cite{petrov2014elastic,goban2018emergence}, where quantum tricritical points naturally occur\,\cite{zwerger2019quantum}. Interestingly, some experimental AMO settings~\cite{defenu2023longrange} can be considered either as isolated microcanonical systems, such as ensembles of dipolar atoms/molecules~\cite{griesmaier2005bose,micheli2006toolbox,ni2008high}, or as canonical systems in 
contact with a thermal bath, such as cold atoms in cavities\,\cite{mivehvar2021cavity}. The latter systems are particularly relevant to our study since the interactions mediated by the cavity photons are global and flat, providing the optimal platform to experimentally verify our findings\,\cite{morrison2008dynamical,larson2010circuit,schutz2016dissipation,keller2018quenches}. In fact, cavity QED experiments were recently employed to investigate the peculiar pre-thermalization dynamics of long-range assemblies\,\cite{wu2023signatures}.
From a broader perspective, the realisation of fully-connected quantum Hamiltonians is also relevant to the optimisation of classical combinatorial problems via adiabatic quantum computing~\cite{albash2018adiabatic}. 

\emph{The model:}
It is convenient to discuss our findings in a concrete example of long-range quantum system, where the extension of the classical picture to the quantum realm can be carried out explicitly. Therefore, we introduce the Hamiltonian of a long-range quantum ferromagnetic spin-1/2 chain with $4$-spin interactions
\begin{align}
\label{qbc_h}
\mathcal{H}=-\frac{J}{N}\left(\sum_{\ell}\sigma^{z}_{\ell}\right)^{2}-h\sum_{\ell}\sigma_{\ell}^{x}
-\frac{K}{N^{3}}\left(\sum_{\ell}\sigma_{\ell}^{z}\right)^{4},
\end{align}
where the summations are taken over all values of the index $\ell\in\{1,\cdots,N\}$ which labels the $N$ sites of the lattice. The $\sigma^{\mu}_{\ell}$ operators are the $\mu=x,y,z$ Pauli matrices at site $\ell$
\begin{equation}
\sigma^x_{\ell}=\begin{pmatrix}
0 & 1\\
1 & 0
\end{pmatrix} \,,\,
\sigma^y_{\ell}=\begin{pmatrix}
0 & -i \\
i & 0 \end{pmatrix} \,,\,
\sigma^z_{\ell}=\begin{pmatrix}
1 & 0\\
0 & -1
\end{pmatrix}.
\end{equation}
In the following, we restrict the discussion to the fully ferromagnetic case $J,K>0$.

Let us define the vector operator
\begin{equation}
\mathbf{S}=\frac{1}{2}\sum_{\ell}\boldsymbol{\sigma}_{\ell}
\end{equation}
where we use the bold face vector notation $\mathbf{S}=(S^x, S^y, S^z)$ and similarly for $\boldsymbol{\sigma}$.
In terms of this operator, the Hamiltonian takes the form
\begin{align}
\mathcal{H}=-\frac{4J}{N}\left(S^z\right)^{2}-2hS^x
-\frac{16K}{N^{3}}(S^z)^4.  
\end{align}
The Hamiltonian in Eq.\,\eqref{qbc_h} reduces to the celebrated Lipkin-Meshkov-Glick (LMG) model in the $K\to0$ limit\,\cite{lipkin1965validity,meshkov1965validity,glick1965validity}. 
There, the system is known to possess a $T=0$ quantum critical point at $h=h_{c}=2 J$,
where a phase transition occurs between a paramagnetic state, fully aligned along $x$, and a ferromagnetic state with a non vanishing magnetization along $z$. 

Models like in Eq.\,\eqref{qbc_h} have been used to study various physical systems in both canonical and microcanonical settings. In the canonical setting, the quantum critical point is closely related to the Dicke model\,\cite{dicke1954coherence}, observable by coupling the motional degrees of freedom of a Bose gas with a cavity's standing wave-field\,\cite{baumann2010dicke,landig2015measuring}. The Dicke model can be mapped onto the LMG model\,\cite{reslen2005direct}, showing that the transition from a disordered atom cloud to a self-organized phase is a second-order phase transition in the same universality class as the Hamiltonian-Mean-Field model\,\cite{campa2014physics,schutz2015thermodynamics}. Spin Hamiltonians like Eq.,\eqref{qbc_h} can also be realized by coupling the internal degrees of freedom of atoms with the cavity field\,\cite{leroux2010implementation,bentsen2019integrable,davis2019photon,davis2020protecting}. In contrast, systems like coupled Bose-Einstein condensates (BECs), the Bose-Hubbard model in a double well potential\,\cite{gallemi2016quantum}, spin-$1$ BECs\,\cite{ho1998spinor,ohmi1998bose,stenger1998spin,chang2004observation,schmaljohann2004dynamics,hoang2016parametric}, or Rydberg atoms in the blockade regime\,\cite{weimer2010rydberg,henkel2010three,gil2014spin,zeiher2015microscopic,jau2016entangling} are better described in a microcanonical setting. Therefore, it would be of interest to study a model like Eq.\eqref{qbc_h} within both canonical and microcanonical settings.

An important aspect of model (\ref{qbc_h}) is the inclusion of multi-spin interactions. In fact, previous studies have been limited to the Hamiltonian\,\eqref{qbc_h} in the $K\to 0$ limit. The study of model (\ref{qbc_h}) fits well within the current experimental endeavours that are pushing toward the quantum control of multi-body interactions\,\cite{will2010time,buchler2007three}. Multi-spin interactions have also been applied to model order-disorder ferroelectric transitions~\cite{Delre}.

\emph{Model analysis:}
Hamiltonian (\ref{qbc_h}) commutes with the total spin operator
\begin{equation}
\mathbf{S}^2=\left(\frac{1}{2}\sum_{\ell}\boldsymbol{\sigma}_{\ell}\right)^2.
\end{equation}
As a result, the Hilbert space decomposes into a set of subspaces, each with a fixed value
of total spin $S=0,\ldots,M$, with $M=N/2$, where for simplicity we restricted the lattice to have an even number of 
sites. The eigenvalue of the total spin operator in the $S$ subspace is $S(S+1)$.
One notes that one has $g(S)$ possible ways to arrange the microscopic $1/2$-spins in order to form a total spin 
$S$, with~\cite{Delre}
\begin{align}
\label{counting}
g(S)={2M\choose M+S}-{2M\choose M+S+1}.
\end{align}
This formula can be verified by observing that the number of states with $S^z=S$ is given by the first
term on the r.h.s. of Eq.~(\ref{counting}). However, some of these states belong to higher total spin $S$ sectors, whose number is given by the second term in the equation above.

We proceed by calculating the free energy and the entropy of the model. We thus define the partition
function $Z$ and the phase space volume $\Omega$ as
\begin{align}
\label{f_en}
Z(\beta,J,h,K)&=\mathrm{Tr}\left[e^{-\beta \mathcal{H}}\right]\\
\label{en}
\Omega(E,J,h,K)&=\mathrm{Tr}\left[\delta(E-\mathcal{H})\right]
\end{align}
where $\delta(\cdots)$ is the Dirac $\delta$-function and $\beta=1/T$ is the inverse temperature.  Using the Hilbert space decomposition and the degeneracy $g(S)$, the traces can be more explicitly expressed as
\begin{align}
\label{f_sum}
&Z(\beta,J,h,K)=\sum_{S}g(S)\sum_{S^z=-S}^{S}\langle S,S^z|e^{-\beta \mathcal{H}}|S,S^z\rangle,\\
\label{en_sum}
&\Omega(E,J,h,K)=\sum_{S}g(S)\sum_{S^z=-S}^{S} \langle S,S^z|\delta(E- \mathcal{H})|S,S^z\rangle.
\end{align}

Due to the mean-field nature of the interaction, the summations in these formulas can be evaluated straightforwardly in the thermodynamic limit. Let us define $S=Ms$ and note that $s$ becomes a continuous variable in the interval $[0,1]$ as $M \to \infty$. Moreover, the energy density is defined as $\varepsilon=E/N$. The magnetization can be written as a classical vector $\mathbf{S}=Ms\mathbf{m}$, where $\mathbf{m} \equiv (m_{x},m_{y},m_{z})=(\sin\theta\cos\phi,\sin\theta\sin\phi,\cos\theta)$ is a unit vector representing the orientation of the magnetization and $s\mathbf{m}$ is the magnetization vector per spin. The sums can thus be replaced by integrals yielding~\cite{Lieb,Granet}
\begin{widetext}
\begin{align}
\label{pf}
& Z(\beta,J,h,K)=\int_{0}^{1} ds \frac{N(Ns+1)}{8\pi} g(Ms) \int e^{-N\beta e(s,\theta,\phi,J,h,K)}\sin\theta d\theta d\phi,\\
&\Omega(\varepsilon,J,h,K)= \int_{0}^{1} ds \frac{N(Ns+1)}{8\pi} g(Ms) \int \delta(E-  N e(s,\theta,\phi,J, h,K))\sin\theta d\theta d\phi.
\end{align}
where
\begin{align}
\label{energy}
e(s,\theta,\phi, J, h, K)=-Js^{2}\cos^{2}\theta-Ks^{4}\cos^{4}\theta-hs\sin\theta\cos\phi.
\end{align}
In the thermodynamic limit one can approximate the $g(Ms)$ factor using Stirling formula, giving the entropy 
\begin{align}
\label{entropy}
&{\cal S}(s)=\frac{\log(g(Ms))}{2M}\approx-\frac{1+s}{2}\log\left(\frac{1+s}{2}\right)-\frac{1-s}{2}\log\left(\frac{1-s}{2}\right).
\end{align}
The partition sum becomes
\begin{align}
\label{z}
&Z(\beta,J,h,K)=\int_{0}^{1}ds\frac{N(Ns+1)}{8\pi}\int e^{-N\beta \left(e(s,\theta,\phi,J,h,K)-{\cal S}(s)/\beta\right)}\sin\theta d\theta d\phi.
\end{align}
The phase-space volume $\Omega(\varepsilon,J,h,K)$ can be calculated using the Fourier representation of the $\delta$-function yielding 
\begin{equation}
\label{o}
\Omega(\varepsilon,J,h,K)=\int_{-\infty}^{+\infty} \frac{d\lambda}{2\pi}\int_{0}^{1} ds\,\frac{N(Ns+1)}
{8\pi}\int e^{-i\lambda N\left(\varepsilon-e(s,\theta,\phi,J,h,K)+{\cal S}(s)/\lambda \right)}\sin\theta d\theta d\phi.
\end{equation}
\end{widetext}

As the thermodynamic limit is approached the integrals in Eqs.\,\eqref{z} and\,\eqref{o} are dominated by the saddle points of the arguments of the exponentials. In both cases, the value of $\phi$ is unambiguously fixed at $\phi=0$, leaving only one single free parameter in the canonical ensemble $m_{z}=\cos\theta\in[-1,1]$. On the other hand, the microcanonical ensemble also requires an additional extremization with respect to the parameter $\lambda$, which results in a constraint on the average energy of the system $\langle\hat{\mathcal{H}}\rangle=\varepsilon$.
In what follows, we first consider the phase diagram in the ground state and then analyse the finite-temperature
phase diagrams in the two ensembles.

\emph{Ground-state phase-diagram:}
Quantum critical behavior occurs at zero temperature $T=0$, where thermal fluctuations do not affect quantum coherence. In this limit, the system configuration matches the Hamiltonian's ground state, so both the canonical and microcanonical ensembles yield the same phase diagram. Considering non-negative parameters  $J,h$ and $K$, the ground state of the model is always in the $s=1$ subset of the spectrum, as shown by the energy expression~\eqref{energy}.
One then needs to minimize the energy with respect to $\theta$. Expressing the energy \eqref{energy} in terms of $m_{z} \equiv \cos\theta$ and expanding it in powers of $m_z$, one obtains
\begin{align}
&\varepsilon(m_{z},J,h,K)=-Jm_{z}^{2}-Km_{z}^{4}-h\sqrt{1-m_{z}^{2}}\nonumber \\
&\approx -h +\left(\frac{h}{2}-J\right)m_{z}^{2}+\left(\frac{h}{8}-K\right)m_{z}^{4}+\frac{h}{16}m_{z}^{6} +O(m_{z}^8).
\end{align}
This energy yields a second order critical line at $h=h_{c}=2J$, separating a disordered state $m_z=0$ from and ordered one
with non-vanishing $m_z$. This result is valid as long as the fourth-order term in the expansion of the energy is positive.
The transition becomes first-order when the fourth-order term changes sign at the tricritical point given by $h/J=2$ and $K/J=1/4$. Close to this point, the first-order transition is given by
\begin{align}
\label{firstorderlinetequalzero}
\frac{K}{J}=\frac{h}{8J}+\frac{1}{2}\sqrt{\frac{h}{J}\frac{(h/J-2)}{2}}   
\end{align}

\begin{figure}
\centering
\includegraphics[width=\linewidth]{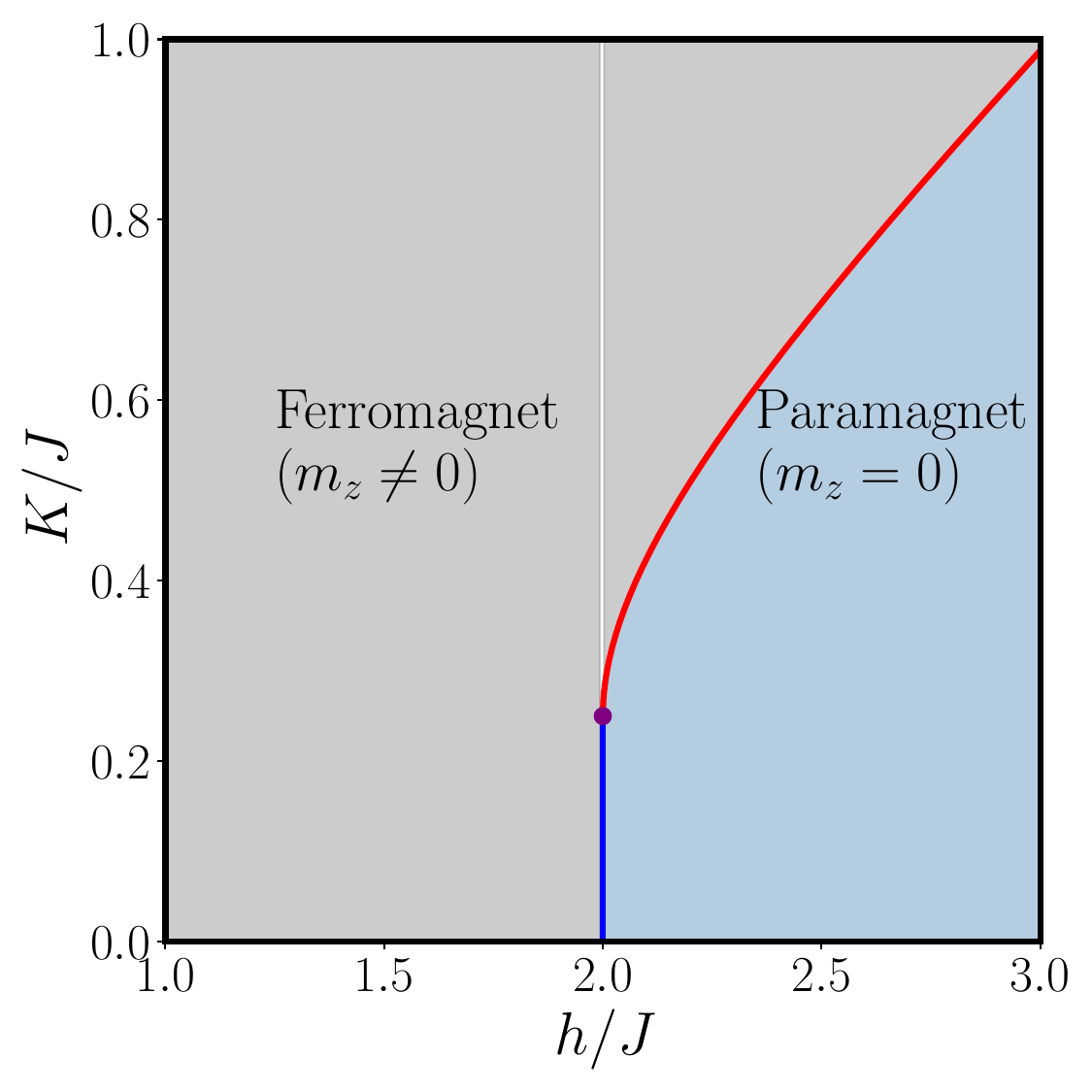}
\caption{\label{Fig1} The ground-state ($T=0$)  phase diagram of the model defined in Eq.\,\eqref{qbc_h} in the $(K/J,h/J)$ plane. It corresponds to both microcanonical and canonical ensembles. The phase-diagram displays a paramagnetic phase with $m_{z}=0$ (blue
shaded area) in the high h/J region. This phase is separated from the ferromagnetic phase where $m_z$ is non vanishing (gray shaded area) by a second order transition line at $h/J=2$ for $K/J<1/4$ (blue line). The transition becomes first-order (red line) at the tricritical point $(K/J=1/4,h/J=2$). The first-order line is given in Eq.~\,\eqref{firstorderlinetequalzero}.}
\end{figure}

The complete ground-state phase-diagram, shared by both the canonical and microcanonical ensembles,  is given in Fig.\ref{Fig1}. In the following, we calculate the phase-diagram at
finite temperature, where we find that the two ensembles yield different phase-diagrams.

\emph{Phase-diagram in the canonical ensemble:}
Let us consider the free energy of the model
\begin{align}
\label{freeenergy}
    &f(\beta,J,h,K)=e-{\cal S}/\beta=-Js^2m_z^2-Ks^4m_z^4\nonumber\\
    &-hs\sqrt{1-m_z^2}+\frac{1}{\beta}
    \left[\frac{1+s}{2}\ln \frac{1+s}{2}+\frac{1-s}{2}\ln \frac{1-s}{2}\right]~.
\end{align}
In order to find the equilibrium state of the system, one needs to minimize $f$ with respect to 
$s$ and $m$. Minimizing \eqref{freeenergy} with respect to $s$ first, we obtain an expansion of $s$ as a function of $m_z$,
\begin{align}
\label{expansionofs}
    s=s_0+am_z^2+O(m_z^4)
\end{align}
where 
\begin{subequations}
\label{coefficients}
\begin{align}
    s_0&=\tanh (\beta h)~,\label{coefficientsa}\\
    a&=\beta (1-s_0^2)(2Js_0-\frac{h}{2})~.\label{coefficientsb}
\end{align}
\end{subequations} 
Inserting the expansion \eqref{expansionofs} in the free energy \eqref{freeenergy}, one obtains an
expansion of $f$ in powers of $m_z^2$:
\begin{align} 
f=f_0+b_2m_z^2+b_4m_z^4+O(m_z^6)~,
\end{align}
with
\begin{subequations}
\begin{align}
\label{coefficientsfree}
    f_0&= -hs_0+\frac{1}{\beta}{\cal S}(s_0)~,\\
    b_2&=\frac{1}{2} h s_0 -J s_0^2~.
\end{align}
\end{subequations}

At criticality $b_2=0$, yielding
\begin{align}
s_0=\frac{h}{2J}~,
\end{align}
and the critical line is given by
\begin{align}
    \frac{h}{2J}=\tanh{\beta h}~,
\end{align}
as long as $b_4>0$. At low temperature the critical line is given, to leading order,
by
\begin{align}
    \frac{h}{2J} \approx 1-2 e^{-2 \beta h}~.
\end{align}
To proceed, we evaluate $b_4$ and locate the tricritical point at $b_2=b_4=0$. We first
expand the entropy in powers $\delta s$ for $s=s_0+\delta s$,
\begin{align}
    {\cal S} (s) \approx {\cal S}(s_0)+\frac{1}{2\beta}\ln \left(\frac{1+s_0}{1-s_0}\right)\delta s
    +\frac{1}{2\beta}\ln \left(\frac{1}{1-s_0^2}\right) \delta s^2 .
\end{align}
Using this expansion with $\delta s=am_z^2$, one finds that on the critical line $b_2=0$ the expression
for $b_4$ is
\begin{align}
    b_4=-(2J s_0-\frac{1}{2}h)a - (K s_0^4 -\frac{1}{8} h s_0) 
    +\frac{1}{2\beta}\frac{1}{1-s_0^2}a^2~.
\end{align}
Note that, due the fact that $\partial f/\partial s|_{s_0}=0$, higher order terms in the expansion
\eqref{expansionofs} of $s$  do not contribute to $b_4$.  Using \eqref{coefficientsb} for $a$, we
obtain
\begin{align}
    b_4= -(Ks_0^4-\frac{h}{8}s_0)-\frac{1}{8} \beta h^2 (1-s_0^2)~,
\end{align}
where at low temperature 
\begin{align}
    s_0=1 -2 e^{-2\beta h}~.
\end{align}
We finally arrive at the following expressions for the critical line ($b_2=0$) and the
tricritical point ($b_2=b_4=0$) in the canonical ensemble ([CE])
\begin{align}
    &b_2=0: \,\,\,\, h_c[CE]=2J(1-2e^{-4\beta J})~, \label{canonicalcriticala}\\
    &b_4=0: \,\,\,\, K_{tcp}[CE]=\frac{J}{4}-2 \beta J^2 e^{-4\beta J} \label{canonicalcriticalb}~.
\end{align}

\emph{Phase-diagram in the microcanonical ensemble:}
The microcanonical phase-diagram can be readily obtained by minimizing the energy at constant entropy. The
energy is given by
\begin{align}
\label{energymicro}
    &\varepsilon=-J s^2 m_z^2 -K s^4 m_z^4 -hs \sqrt{1-m_z^2} \nonumber\\
    &\approx -hs +\left(\frac{1}{2}hs
    -Js^2\right) m_z^2 +\left(\frac{1}{8} hs -Ks^4 \right) m_z^4 +O(m_z^6)~.
\end{align}
The entropy ${\cal S}(s)$ \eqref{entropy} is a function of $s$ only and thus one has to minimize the energy $\varepsilon$
with respect to $m_z$ at fixed $s$. The resulting critical line is
\begin{align}
\label{criticalline}
    \frac{1}{2} hs-Js^2=0~,
\end{align}
which, together with 
\begin{align}
\label{condition}
    \frac{1}{8} hs -K s^4=0~,
\end{align}
yields the tricritical point. To proceed, one has to express $s$ in terms of the temperature. On the critical line, where $m_z=0$,
the energy is given by $\varepsilon=-hs$. Thus,
\begin{align}
    \beta=\frac{\partial {\cal S}}{\partial \varepsilon}=-\frac{1}{h} \frac{\partial {\cal S}}{\partial s}=
    \frac{1}{2h} \ln \frac{1-s}{1+s}~,
\end{align}
which gives
\begin{align}
\label{es}
    s=\tanh{\beta h} \approx 1 -2 e^{-2\beta h}~.
\end{align}
Inserting expression \eqref{es} in \eqref{criticalline}, the microcanonical ([MCE]) critical line becomes
\begin{align}
\label{microcritical}
    h_c[MCE]=2J \left( 1-2e^{-4 \beta J} \right)~.
\end{align}
On the critical line, Eq. \eqref{condition} becomes $J=4Ks^2$, which yields the tricritical point at
\begin{align}
\label{microtricritical}
    K_{tcp}[MCE]=J\left(\frac{1}{4} + e^{-4\beta J}\right)~.
\end{align}

\begin{figure}
\centering
\includegraphics[width=\linewidth]{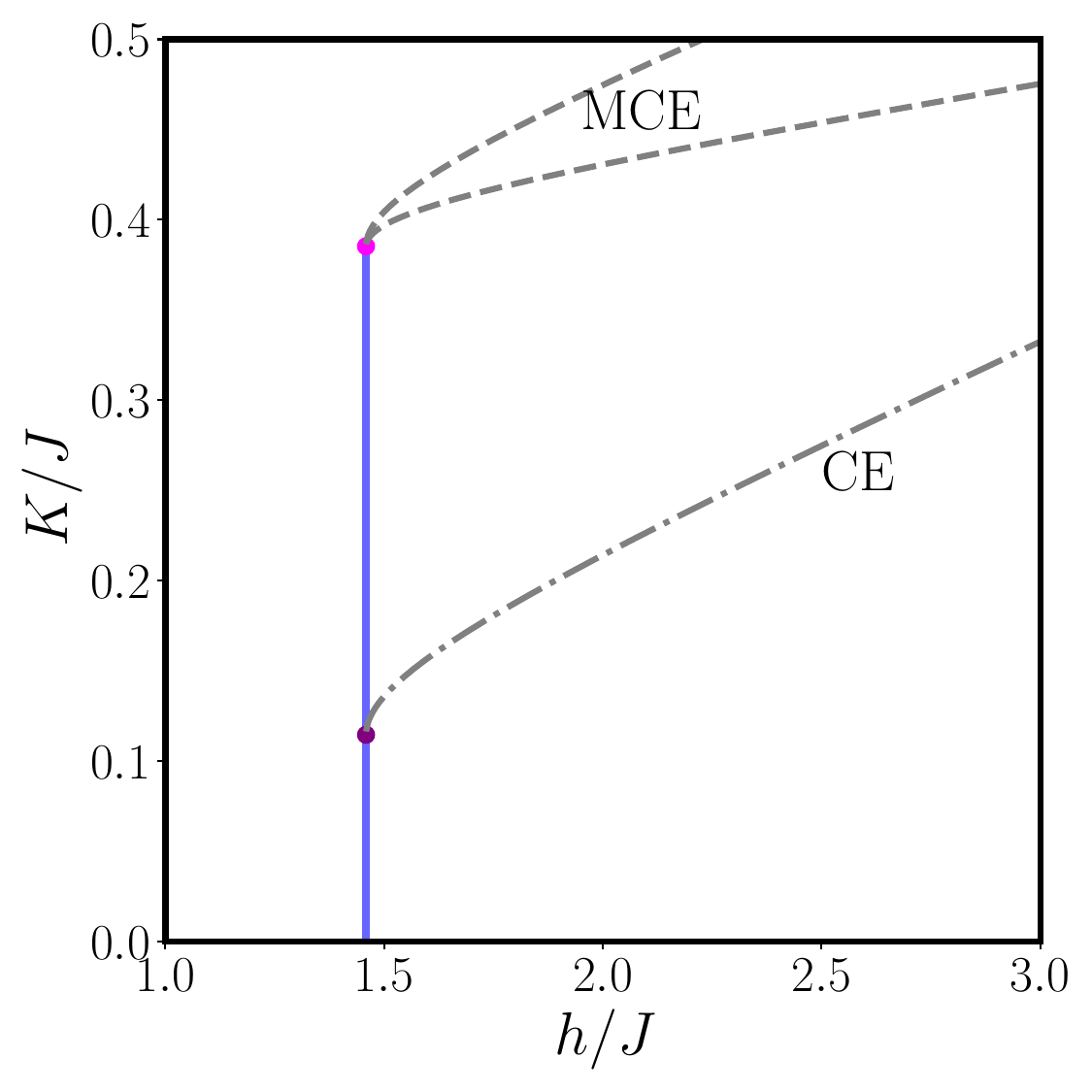}\caption{\label{Fig2} 
The canonical and microcanonical $(h/J,K/J)$ phase-diagrams at a given temperature ($\beta J=1/2$). The microcanonical critical line coincides with the canonical one, but extends beyond the canonical
tricritical point. While the critical line (blue) is drawn in scale, the first order lines (gray dashed lines in the microcanonical ensemble and a gray dot-dashed line in the canonical one) are only drawn schematically.}
\end{figure}

\begin{figure}
\centering
\includegraphics[width=\linewidth]{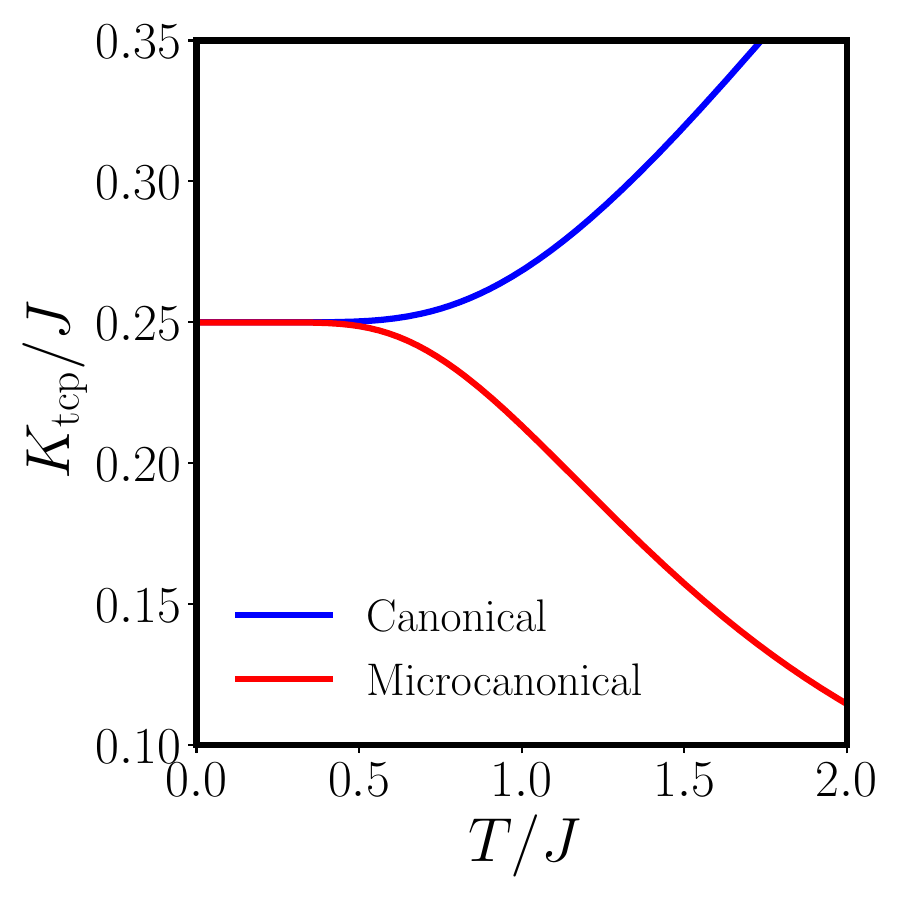}\caption{\label{Fig3}
The position of the trictitical point $K/J$ against $T/J$ in the canonical and microcanonical ensembles.}
\end{figure}

When compared with the canonical analysis, this result presents an example of ensemble inequivalence. 
While the two ensembles lead to the same expression for the critical lines, \eqref{canonicalcriticala}, \eqref{microcritical}, they display distinct tricritical points. At a given temperature, the canonical tricritical point \eqref{canonicalcriticalb} is located at a lower value of $K/J$ than the microcanonical one \eqref{microtricritical}, (see Fig.\ref{Fig2} for the $(h/J,K/J)$ phase-diagram at a given low temperature).
In Fig.\ref{Fig3} we display the tricritical coupling $K/J$ in the two ensembles, \eqref{canonicalcriticalb} and \eqref{microtricritical}, as a function of $T/J$ at low temperatures. While the tricritical points coincide at $T=0$, the microcanonical one changes slower with temperature. Note that at any given temperature the magnetic field $h/J$ at the tricritical point is the same in the two ensembles.

\emph{Conclusions:} In this paper, we studied the phase diagram of a model with long-range and multi-spin interactions, which exhibits a paramagnetic to ferromagnetic quantum phase transition at zero temperature. This transition has first-order and second-order branches separated by a tricritical point. At finite temperature, we showed that the model exhibits different phase diagrams in the canonical and microcanonical ensembles. While the two ensembles yield the same phase diagram at 
$T=0$, they differ at finite temperatures. Notably, the position of the tricritical point varies between the ensembles, with the finite temperature correction being larger in the canonical ensemble than in the microcanonical ensemble.
Thus, the quantum tricritical points of long-range systems split due to finite temperature corrections, an effect that could not be predicted by the quantum-to-classical correspondence. 

As AMO techniques continue to advance, our findings will become crucial to describe the critical scaling region of experimental platforms, where quantum fluctuations compete with long-range interactions. Indeed, while the Hamiltonian in Eq.\,\eqref{pf} with $K=0$ has already been employed in the description of cavity QED platforms\,\cite{morrison2008dynamical,morrison2008collective,cosme2023bridging}, the realization of the four body term at $K>0$ may be achieved by exploiting recent finding on cavity-mediated pair creation\,\cite{finger2023spin}.

On general grounds, ensemble inequivalence is expected to occur whenever the canonical transition becomes first order in long-range quantum systems. Future investigations shall clarify how the phenomenon quantitatively arises in the case the first term in Hamiltonian\,\eqref{qbc_h} decays as a power-law of the distance $r^{-\alpha}$ with $\alpha<1$.

\emph{Acknowledgments:} Valuable discussions with G. Gori and A. Trombettoni are gratefully acknowledged. Relevant discussions on the experimental feasibility of our research with T. Donner are also acknowledged. This research was funded by the Swiss National Science Foundation (SNSF) grant number 200021 207537, by the Deutsche Forschungsgemeinschaft (DFG, German Research Foundation) under Germany's Excellence Strategy EXC2181/1-390900948 (the Heidelberg STRUCTURES Excellence Cluster) and the Swiss State Secretariat for Education, Research and Innovation (SERI). This work is part of MUR-PRIN2017 project “Coarse-grained description for non-equilibrium systems and transport phenomena (CONEST)" No. 201798CZL whose partial financial support is acknowledged. DM acknowledges the support of the Center for Scientific
Excellence of the Weizmann Institute of Science. This research was supported in part by grants NSF PHY-1748958 and PHY-2309135 to the Kavli Institute for Theoretical Physics (KITP).

\bibliography{main}

\begin{thebibliography}{71}%
\makeatletter
\providecommand \@ifxundefined [1]{%
 \@ifx{#1\undefined}
}%
\providecommand \@ifnum [1]{%
 \ifnum #1\expandafter \@firstoftwo
 \else \expandafter \@secondoftwo
 \fi
}%
\providecommand \@ifx [1]{%
 \ifx #1\expandafter \@firstoftwo
 \else \expandafter \@secondoftwo
 \fi
}%
\providecommand \natexlab [1]{#1}%
\providecommand \enquote  [1]{``#1''}%
\providecommand \bibnamefont  [1]{#1}%
\providecommand \bibfnamefont [1]{#1}%
\providecommand \citenamefont [1]{#1}%
\providecommand \href@noop [0]{\@secondoftwo}%
\providecommand \href [0]{\begingroup \@sanitize@url \@href}%
\providecommand \@href[1]{\@@startlink{#1}\@@href}%
\providecommand \@@href[1]{\endgroup#1\@@endlink}%
\providecommand \@sanitize@url [0]{\catcode `\\12\catcode `\$12\catcode
  `\&12\catcode `\#12\catcode `\^12\catcode `\_12\catcode `\%12\relax}%
\providecommand \@@startlink[1]{}%
\providecommand \@@endlink[0]{}%
\providecommand \url  [0]{\begingroup\@sanitize@url \@url }%
\providecommand \@url [1]{\endgroup\@href {#1}{\urlprefix }}%
\providecommand \urlprefix  [0]{URL }%
\providecommand \Eprint [0]{\href }%
\providecommand \doibase [0]{https://doi.org/}%
\providecommand \selectlanguage [0]{\@gobble}%
\providecommand \bibinfo  [0]{\@secondoftwo}%
\providecommand \bibfield  [0]{\@secondoftwo}%
\providecommand \translation [1]{[#1]}%
\providecommand \BibitemOpen [0]{}%
\providecommand \bibitemStop [0]{}%
\providecommand \bibitemNoStop [0]{.\EOS\space}%
\providecommand \EOS [0]{\spacefactor3000\relax}%
\providecommand \BibitemShut  [1]{\csname bibitem#1\endcsname}%
\let\auto@bib@innerbib\@empty
\bibitem [{\citenamefont {Lukin}(2003)}]{lukin2003trapping}%
  \BibitemOpen
  \bibfield  {author} {\bibinfo {author} {\bibfnamefont {M.~D.}\ \bibnamefont
  {Lukin}},\ }\bibfield  {title} {\bibinfo {title} {Colloquium: Trapping and
  manipulating photon states in atomic ensembles},\ }\href
  {https://doi.org/10.1103/RevModPhys.75.457} {\bibfield  {journal} {\bibinfo
  {journal} {Rev. Mod. Phys.}\ }\textbf {\bibinfo {volume} {75}},\ \bibinfo
  {pages} {457} (\bibinfo {year} {2003})}\BibitemShut {NoStop}%
\bibitem [{\citenamefont {{Saffman}}\ \emph {et~al.}(2010)\citenamefont
  {{Saffman}}, \citenamefont {{Walker}},\ and\ \citenamefont
  {{M{\o}lmer}}}]{saffman2010quantum}%
  \BibitemOpen
  \bibfield  {author} {\bibinfo {author} {\bibfnamefont {M.}~\bibnamefont
  {{Saffman}}}, \bibinfo {author} {\bibfnamefont {T.~G.}\ \bibnamefont
  {{Walker}}},\ and\ \bibinfo {author} {\bibfnamefont {K.}~\bibnamefont
  {{M{\o}lmer}}},\ }\bibfield  {title} {\bibinfo {title} {{Quantum information
  with Rydberg atoms}},\ }\href@noop {} {\bibfield  {journal} {\bibinfo
  {journal} {Rev. Mod. Phys.}\ }\textbf {\bibinfo {volume} {82}},\ \bibinfo
  {pages} {2313} (\bibinfo {year} {2010})}\BibitemShut {NoStop}%
\bibitem [{\citenamefont {Britton}\ \emph {et~al.}(2012)\citenamefont
  {Britton}, \citenamefont {Sawyer}, \citenamefont {Keith}, \citenamefont
  {Wang}, \citenamefont {Freericks}, \citenamefont {Uys}, \citenamefont
  {Biercuk},\ and\ \citenamefont {Bollinger}}]{britton2012engineered}%
  \BibitemOpen
  \bibfield  {author} {\bibinfo {author} {\bibfnamefont {J.~W.}\ \bibnamefont
  {Britton}}, \bibinfo {author} {\bibfnamefont {B.~C.}\ \bibnamefont {Sawyer}},
  \bibinfo {author} {\bibfnamefont {A.~C.}\ \bibnamefont {Keith}}, \bibinfo
  {author} {\bibfnamefont {C.~C.~J.}\ \bibnamefont {Wang}}, \bibinfo {author}
  {\bibfnamefont {J.~K.}\ \bibnamefont {Freericks}}, \bibinfo {author}
  {\bibfnamefont {H.}~\bibnamefont {Uys}}, \bibinfo {author} {\bibfnamefont
  {M.~J.}\ \bibnamefont {Biercuk}},\ and\ \bibinfo {author} {\bibfnamefont
  {J.~J.}\ \bibnamefont {Bollinger}},\ }\bibfield  {title} {\bibinfo {title}
  {{Engineered two-dimensional Ising interactions in a trapped-ion quantum
  simulator with hundreds of spins}},\ }\href@noop {} {\bibfield  {journal}
  {\bibinfo  {journal} {Nature}\ }\textbf {\bibinfo {volume} {484}},\ \bibinfo
  {pages} {489} (\bibinfo {year} {2012})}\BibitemShut {NoStop}%
\bibitem [{\citenamefont {Bloch}\ \emph {et~al.}(2008)\citenamefont {Bloch},
  \citenamefont {Dalibard},\ and\ \citenamefont {Zwerger}}]{bloch2008many}%
  \BibitemOpen
  \bibfield  {author} {\bibinfo {author} {\bibfnamefont {I.}~\bibnamefont
  {Bloch}}, \bibinfo {author} {\bibfnamefont {J.}~\bibnamefont {Dalibard}},\
  and\ \bibinfo {author} {\bibfnamefont {W.}~\bibnamefont {Zwerger}},\
  }\bibfield  {title} {\bibinfo {title} {{Many-body physics with ultracold
  gases}},\ }\href {https://doi.org/10.1103/RevModPhys.80.885} {\bibfield
  {journal} {\bibinfo  {journal} {Reviews of Modern Physics}\ }\textbf
  {\bibinfo {volume} {80}},\ \bibinfo {pages} {885} (\bibinfo {year}
  {2008})}\BibitemShut {NoStop}%
\bibitem [{\citenamefont {Blatt}\ and\ \citenamefont
  {Roos}(2012)}]{blatt2012quantum}%
  \BibitemOpen
  \bibfield  {author} {\bibinfo {author} {\bibfnamefont {R.}~\bibnamefont
  {Blatt}}\ and\ \bibinfo {author} {\bibfnamefont {C.~F.}\ \bibnamefont
  {Roos}},\ }\bibfield  {title} {\bibinfo {title} {{Quantum simulations with
  trapped ions}},\ }\href@noop {} {\bibfield  {journal} {\bibinfo  {journal}
  {Nat. Phys.}\ }\textbf {\bibinfo {volume} {8}},\ \bibinfo {pages} {277}
  (\bibinfo {year} {2012})}\BibitemShut {NoStop}%
\bibitem [{\citenamefont {Monroe}\ \emph {et~al.}(2021)\citenamefont {Monroe},
  \citenamefont {Campbell}, \citenamefont {Duan}, \citenamefont {Gong},
  \citenamefont {Gorshkov}, \citenamefont {Hess}, \citenamefont {Islam},
  \citenamefont {Kim}, \citenamefont {Linke}, \citenamefont {Pagano},
  \citenamefont {Richerme}, \citenamefont {Senko},\ and\ \citenamefont
  {Yao}}]{monroe2021programmable}%
  \BibitemOpen
  \bibfield  {author} {\bibinfo {author} {\bibfnamefont {C.}~\bibnamefont
  {Monroe}}, \bibinfo {author} {\bibfnamefont {W.~C.}\ \bibnamefont
  {Campbell}}, \bibinfo {author} {\bibfnamefont {L.-M.}\ \bibnamefont {Duan}},
  \bibinfo {author} {\bibfnamefont {Z.-X.}\ \bibnamefont {Gong}}, \bibinfo
  {author} {\bibfnamefont {A.~V.}\ \bibnamefont {Gorshkov}}, \bibinfo {author}
  {\bibfnamefont {P.~W.}\ \bibnamefont {Hess}}, \bibinfo {author}
  {\bibfnamefont {R.}~\bibnamefont {Islam}}, \bibinfo {author} {\bibfnamefont
  {K.}~\bibnamefont {Kim}}, \bibinfo {author} {\bibfnamefont {N.~M.}\
  \bibnamefont {Linke}}, \bibinfo {author} {\bibfnamefont {G.}~\bibnamefont
  {Pagano}}, \bibinfo {author} {\bibfnamefont {P.}~\bibnamefont {Richerme}},
  \bibinfo {author} {\bibfnamefont {C.}~\bibnamefont {Senko}},\ and\ \bibinfo
  {author} {\bibfnamefont {N.~Y.}\ \bibnamefont {Yao}},\ }\bibfield  {title}
  {\bibinfo {title} {Programmable quantum simulations of spin systems with
  trapped ions},\ }\href {https://doi.org/10.1103/RevModPhys.93.025001}
  {\bibfield  {journal} {\bibinfo  {journal} {Rev. Mod. Phys.}\ }\textbf
  {\bibinfo {volume} {93}},\ \bibinfo {pages} {025001} (\bibinfo {year}
  {2021})}\BibitemShut {NoStop}%
\bibitem [{\citenamefont {Mivehvar}\ \emph {et~al.}(2021)\citenamefont
  {Mivehvar}, \citenamefont {Piazza}, \citenamefont {Donner},\ and\
  \citenamefont {Ritsch}}]{mivehvar2021cavity}%
  \BibitemOpen
  \bibfield  {author} {\bibinfo {author} {\bibfnamefont {F.}~\bibnamefont
  {Mivehvar}}, \bibinfo {author} {\bibfnamefont {F.}~\bibnamefont {Piazza}},
  \bibinfo {author} {\bibfnamefont {T.}~\bibnamefont {Donner}},\ and\ \bibinfo
  {author} {\bibfnamefont {H.}~\bibnamefont {Ritsch}},\ }\bibfield  {title}
  {\bibinfo {title} {Cavity qed with quantum gases: new paradigms in many-body
  physics},\ }\href {https://doi.org/10.1080/00018732.2021.1969727} {\bibfield
  {journal} {\bibinfo  {journal} {Advances in Physics}\ }\textbf {\bibinfo
  {volume} {70}},\ \bibinfo {pages} {1–153} (\bibinfo {year}
  {2021})}\BibitemShut {NoStop}%
\bibitem [{\citenamefont {Defenu}\ \emph {et~al.}(2023)\citenamefont {Defenu},
  \citenamefont {Donner}, \citenamefont {Macr\`{\i}}, \citenamefont {Pagano},
  \citenamefont {Ruffo},\ and\ \citenamefont
  {Trombettoni}}]{defenu2023longrange}%
  \BibitemOpen
  \bibfield  {author} {\bibinfo {author} {\bibfnamefont {N.}~\bibnamefont
  {Defenu}}, \bibinfo {author} {\bibfnamefont {T.}~\bibnamefont {Donner}},
  \bibinfo {author} {\bibfnamefont {T.}~\bibnamefont {Macr\`{\i}}}, \bibinfo
  {author} {\bibfnamefont {G.}~\bibnamefont {Pagano}}, \bibinfo {author}
  {\bibfnamefont {S.}~\bibnamefont {Ruffo}},\ and\ \bibinfo {author}
  {\bibfnamefont {A.}~\bibnamefont {Trombettoni}},\ }\bibfield  {title}
  {\bibinfo {title} {Long-range interacting quantum systems},\ }\href
  {https://doi.org/10.1103/RevModPhys.95.035002} {\bibfield  {journal}
  {\bibinfo  {journal} {Rev. Mod. Phys.}\ }\textbf {\bibinfo {volume} {95}},\
  \bibinfo {pages} {035002} (\bibinfo {year} {2023})}\BibitemShut {NoStop}%
\bibitem [{\citenamefont {Angelini}\ \emph {et~al.}(2014)\citenamefont
  {Angelini}, \citenamefont {Parisi},\ and\ \citenamefont
  {Ricci-Tersenghi}}]{angelini2014relations}%
  \BibitemOpen
  \bibfield  {author} {\bibinfo {author} {\bibfnamefont {M.~C.}\ \bibnamefont
  {Angelini}}, \bibinfo {author} {\bibfnamefont {G.}~\bibnamefont {Parisi}},\
  and\ \bibinfo {author} {\bibfnamefont {F.}~\bibnamefont {Ricci-Tersenghi}},\
  }\bibfield  {title} {\bibinfo {title} {Relations between short-range and
  long-range ising models},\ }\href
  {https://doi.org/10.1103/PhysRevE.89.062120} {\bibfield  {journal} {\bibinfo
  {journal} {Phys. Rev. E}\ }\textbf {\bibinfo {volume} {89}},\ \bibinfo
  {pages} {062120} (\bibinfo {year} {2014})}\BibitemShut {NoStop}%
\bibitem [{\citenamefont {Defenu}\ \emph {et~al.}(2015)\citenamefont {Defenu},
  \citenamefont {Trombettoni},\ and\ \citenamefont
  {Codello}}]{defenu2015fixed}%
  \BibitemOpen
  \bibfield  {author} {\bibinfo {author} {\bibfnamefont {N.}~\bibnamefont
  {Defenu}}, \bibinfo {author} {\bibfnamefont {A.}~\bibnamefont
  {Trombettoni}},\ and\ \bibinfo {author} {\bibfnamefont {A.}~\bibnamefont
  {Codello}},\ }\bibfield  {title} {\bibinfo {title} {{Fixed-point structure
  and effective fractional dimensionality for O( N) models with long-range
  interactions}},\ }\href@noop {} {\bibfield  {journal} {\bibinfo  {journal}
  {Phys. Rev. E}\ }\textbf {\bibinfo {volume} {92}},\ \bibinfo {pages} {052113}
  (\bibinfo {year} {2015})}\BibitemShut {NoStop}%
\bibitem [{\citenamefont {Defenu}\ \emph {et~al.}(2016)\citenamefont {Defenu},
  \citenamefont {Trombettoni},\ and\ \citenamefont
  {Ruffo}}]{defenu2016anisotropic}%
  \BibitemOpen
  \bibfield  {author} {\bibinfo {author} {\bibfnamefont {N.}~\bibnamefont
  {Defenu}}, \bibinfo {author} {\bibfnamefont {A.}~\bibnamefont
  {Trombettoni}},\ and\ \bibinfo {author} {\bibfnamefont {S.}~\bibnamefont
  {Ruffo}},\ }\bibfield  {title} {\bibinfo {title} {Anisotropic long-range spin
  systems},\ }\href {https://doi.org/10.1103/PhysRevB.94.224411} {\bibfield
  {journal} {\bibinfo  {journal} {Phys. Rev. B}\ }\textbf {\bibinfo {volume}
  {94}},\ \bibinfo {pages} {224411} (\bibinfo {year} {2016})}\BibitemShut
  {NoStop}%
\bibitem [{\citenamefont {Defenu}\ \emph {et~al.}(2017)\citenamefont {Defenu},
  \citenamefont {Trombettoni},\ and\ \citenamefont
  {Ruffo}}]{defenu2017criticality}%
  \BibitemOpen
  \bibfield  {author} {\bibinfo {author} {\bibfnamefont {N.}~\bibnamefont
  {Defenu}}, \bibinfo {author} {\bibfnamefont {A.}~\bibnamefont
  {Trombettoni}},\ and\ \bibinfo {author} {\bibfnamefont {S.}~\bibnamefont
  {Ruffo}},\ }\bibfield  {title} {\bibinfo {title} {{Criticality and phase
  diagram of quantum long-range O( N) models}},\ }\href@noop {} {\bibfield
  {journal} {\bibinfo  {journal} {Phys. Rev. B}\ }\textbf {\bibinfo {volume}
  {96}},\ \bibinfo {pages} {1} (\bibinfo {year} {2017})}\BibitemShut {NoStop}%
\bibitem [{\citenamefont {{Defenu}}\ \emph {et~al.}(2020)\citenamefont
  {{Defenu}}, \citenamefont {{Codello}}, \citenamefont {{Ruffo}},\ and\
  \citenamefont {{Trombettoni}}}]{defenu2020criticality}%
  \BibitemOpen
  \bibfield  {author} {\bibinfo {author} {\bibfnamefont {N.}~\bibnamefont
  {{Defenu}}}, \bibinfo {author} {\bibfnamefont {A.}~\bibnamefont {{Codello}}},
  \bibinfo {author} {\bibfnamefont {S.}~\bibnamefont {{Ruffo}}},\ and\ \bibinfo
  {author} {\bibfnamefont {A.}~\bibnamefont {{Trombettoni}}},\ }\bibfield
  {title} {\bibinfo {title} {{Criticality of spin systems with weak long-range
  interactions}},\ }\href {https://doi.org/10.1088/1751-8121/ab6a6c} {\bibfield
   {journal} {\bibinfo  {journal} {Journal of Physics A Mathematical General}\
  }\textbf {\bibinfo {volume} {53}},\ \bibinfo {eid} {143001} (\bibinfo {year}
  {2020})}\BibitemShut {NoStop}%
\bibitem [{\citenamefont {Dauxois}\ \emph {et~al.}(2002)\citenamefont
  {Dauxois}, \citenamefont {Latora}, \citenamefont {Rapisarda}, \citenamefont
  {Ruffo},\ and\ \citenamefont {Torcini}}]{dauxois2002hamiltonian}%
  \BibitemOpen
  \bibfield  {author} {\bibinfo {author} {\bibfnamefont {T.}~\bibnamefont
  {Dauxois}}, \bibinfo {author} {\bibfnamefont {V.}~\bibnamefont {Latora}},
  \bibinfo {author} {\bibfnamefont {A.}~\bibnamefont {Rapisarda}}, \bibinfo
  {author} {\bibfnamefont {S.}~\bibnamefont {Ruffo}},\ and\ \bibinfo {author}
  {\bibfnamefont {A.}~\bibnamefont {Torcini}},\ }\href
  {https://doi.org/10.1007/3-540-45835-2_16} {\emph {\bibinfo {title} {Dynamics
  and Thermodynamics of Systems with Long-Range Interactions}}},\ edited by\
  \bibinfo {editor} {\bibfnamefont {T.}~\bibnamefont {Dauxois}}, \bibinfo
  {editor} {\bibfnamefont {S.}~\bibnamefont {Ruffo}}, \bibinfo {editor}
  {\bibfnamefont {E.}~\bibnamefont {Arimondo}},\ and\ \bibinfo {editor}
  {\bibfnamefont {M.}~\bibnamefont {Wilkens}}\ (\bibinfo  {publisher} {Springer
  Berlin Heidelberg},\ \bibinfo {address} {Berlin, Heidelberg},\ \bibinfo
  {year} {2002})\ pp.\ \bibinfo {pages} {458--487}\BibitemShut {NoStop}%
\bibitem [{\citenamefont {{Campa}}\ \emph {et~al.}(2009)\citenamefont
  {{Campa}}, \citenamefont {{Dauxois}},\ and\ \citenamefont
  {{Ruffo}}}]{campa2009statistical}%
  \BibitemOpen
  \bibfield  {author} {\bibinfo {author} {\bibfnamefont {A.}~\bibnamefont
  {{Campa}}}, \bibinfo {author} {\bibfnamefont {T.}~\bibnamefont {{Dauxois}}},\
  and\ \bibinfo {author} {\bibfnamefont {S.}~\bibnamefont {{Ruffo}}},\
  }\bibfield  {title} {\bibinfo {title} {{Statistical mechanics and dynamics of
  solvable models with long-range interactions}},\ }\href
  {https://doi.org/10.1016/j.physrep.2009.07.001} {\bibfield  {journal}
  {\bibinfo  {journal} {Phys. Rep.}\ }\textbf {\bibinfo {volume} {480}},\
  \bibinfo {pages} {57} (\bibinfo {year} {2009})}\BibitemShut {NoStop}%
\bibitem [{\citenamefont {Campa}\ \emph {et~al.}(2014)\citenamefont {Campa},
  \citenamefont {Dauxois}, \citenamefont {Fanelli},\ and\ \citenamefont
  {Ruffo}}]{campa2014physics}%
  \BibitemOpen
  \bibfield  {author} {\bibinfo {author} {\bibfnamefont {A.}~\bibnamefont
  {Campa}}, \bibinfo {author} {\bibfnamefont {T.}~\bibnamefont {Dauxois}},
  \bibinfo {author} {\bibfnamefont {D.}~\bibnamefont {Fanelli}},\ and\ \bibinfo
  {author} {\bibfnamefont {S.}~\bibnamefont {Ruffo}},\ }\href@noop {} {\emph
  {\bibinfo {title} {{Physics of Long-Range Interacting Systems}}}}\ (\bibinfo
  {publisher} {Oxford Univ. Press},\ \bibinfo {year} {2014})\BibitemShut
  {NoStop}%
\bibitem [{\citenamefont {Levin}\ \emph {et~al.}(2014)\citenamefont {Levin},
  \citenamefont {Pakter}, \citenamefont {Rizzato}, \citenamefont {Teles},\ and\
  \citenamefont {Benetti}}]{levin2014nonequilibrium}%
  \BibitemOpen
  \bibfield  {author} {\bibinfo {author} {\bibfnamefont {Y.}~\bibnamefont
  {Levin}}, \bibinfo {author} {\bibfnamefont {R.}~\bibnamefont {Pakter}},
  \bibinfo {author} {\bibfnamefont {F.~B.}\ \bibnamefont {Rizzato}}, \bibinfo
  {author} {\bibfnamefont {T.~N.}\ \bibnamefont {Teles}},\ and\ \bibinfo
  {author} {\bibfnamefont {F.~P.}\ \bibnamefont {Benetti}},\ }\bibfield
  {title} {\bibinfo {title} {{Nonequilibrium statistical mechanics of systems
  with long-range interactions}},\ }\href
  {https://doi.org/10.1016/j.physrep.2013.10.001} {\bibfield  {journal}
  {\bibinfo  {journal} {Physics Reports}\ }\textbf {\bibinfo {volume} {535}},\
  \bibinfo {pages} {1} (\bibinfo {year} {2014})}\BibitemShut {NoStop}%
\bibitem [{\citenamefont {Barr\'e}\ \emph {et~al.}(2001)\citenamefont
  {Barr\'e}, \citenamefont {Mukamel},\ and\ \citenamefont
  {Ruffo}}]{barre2001inequivalence}%
  \BibitemOpen
  \bibfield  {author} {\bibinfo {author} {\bibfnamefont {J.}~\bibnamefont
  {Barr\'e}}, \bibinfo {author} {\bibfnamefont {D.}~\bibnamefont {Mukamel}},\
  and\ \bibinfo {author} {\bibfnamefont {S.}~\bibnamefont {Ruffo}},\ }\bibfield
   {title} {\bibinfo {title} {Inequivalence of ensembles in a system with
  long-range interactions},\ }\href
  {https://doi.org/10.1103/PhysRevLett.87.030601} {\bibfield  {journal}
  {\bibinfo  {journal} {Phys. Rev. Lett.}\ }\textbf {\bibinfo {volume} {87}},\
  \bibinfo {pages} {030601} (\bibinfo {year} {2001})}\BibitemShut {NoStop}%
\bibitem [{\citenamefont {Kastner}(2011)}]{kastner2011diverging}%
  \BibitemOpen
  \bibfield  {author} {\bibinfo {author} {\bibfnamefont {M.}~\bibnamefont
  {Kastner}},\ }\bibfield  {title} {\bibinfo {title} {{Diverging Equilibration
  Times in Long-Range Quantum Spin Models}},\ }\href@noop {} {\bibfield
  {journal} {\bibinfo  {journal} {Phys. Rev. Lett.}\ }\textbf {\bibinfo
  {volume} {106}},\ \bibinfo {pages} {130601} (\bibinfo {year}
  {2011})}\BibitemShut {NoStop}%
\bibitem [{\citenamefont {Sch\"utz}\ and\ \citenamefont
  {Morigi}(2014)}]{schutz2014prethermalization}%
  \BibitemOpen
  \bibfield  {author} {\bibinfo {author} {\bibfnamefont {S.}~\bibnamefont
  {Sch\"utz}}\ and\ \bibinfo {author} {\bibfnamefont {G.}~\bibnamefont
  {Morigi}},\ }\bibfield  {title} {\bibinfo {title} {Prethermalization of atoms
  due to photon-mediated long-range interactions},\ }\href
  {https://doi.org/10.1103/PhysRevLett.113.203002} {\bibfield  {journal}
  {\bibinfo  {journal} {Phys. Rev. Lett.}\ }\textbf {\bibinfo {volume} {113}},\
  \bibinfo {pages} {203002} (\bibinfo {year} {2014})}\BibitemShut {NoStop}%
\bibitem [{\citenamefont {Sch{\"u}tz}\ \emph {et~al.}(2016)\citenamefont
  {Sch{\"u}tz}, \citenamefont {J{\"a}ger},\ and\ \citenamefont
  {Morigi}}]{schutz2016dissipation}%
  \BibitemOpen
  \bibfield  {author} {\bibinfo {author} {\bibfnamefont {S.}~\bibnamefont
  {Sch{\"u}tz}}, \bibinfo {author} {\bibfnamefont {S.~B.}\ \bibnamefont
  {J{\"a}ger}},\ and\ \bibinfo {author} {\bibfnamefont {G.}~\bibnamefont
  {Morigi}},\ }\bibfield  {title} {\bibinfo {title} {{Dissipation-Assisted
  Prethermalization in Long-Range Interacting Atomic Ensembles}},\ }\href@noop
  {} {\bibfield  {journal} {\bibinfo  {journal} {Phys. Rev. Lett.}\ }\textbf
  {\bibinfo {volume} {117}},\ \bibinfo {pages} {083001} (\bibinfo {year}
  {2016})}\BibitemShut {NoStop}%
\bibitem [{\citenamefont {Mori}(2019)}]{mori2019prethermalization}%
  \BibitemOpen
  \bibfield  {author} {\bibinfo {author} {\bibfnamefont {T.}~\bibnamefont
  {Mori}},\ }\bibfield  {title} {\bibinfo {title} {Prethermalization in the
  transverse-field ising chain with long-range interactions},\ }\href
  {https://doi.org/10.1088/1751-8121/aaf9db} {\bibfield  {journal} {\bibinfo
  {journal} {Journal of Physics A: Mathematical and Theoretical}\ }\textbf
  {\bibinfo {volume} {52}},\ \bibinfo {pages} {054001} (\bibinfo {year}
  {2019})}\BibitemShut {NoStop}%
\bibitem [{\citenamefont {{Defenu}}(2021)}]{defenu2021metastability}%
  \BibitemOpen
  \bibfield  {author} {\bibinfo {author} {\bibfnamefont {N.}~\bibnamefont
  {{Defenu}}},\ }\bibfield  {title} {\bibinfo {title} {{Metastability and
  discrete spectrum of long-range systems}},\ }\href
  {https://doi.org/10.1073/pnas.2101785118} {\bibfield  {journal} {\bibinfo
  {journal} {Proceedings of the National Academy of Science}\ }\textbf
  {\bibinfo {volume} {118}},\ \bibinfo {eid} {e2101785118} (\bibinfo {year}
  {2021})}\BibitemShut {NoStop}%
\bibitem [{\citenamefont {Mukamel}\ \emph {et~al.}(2005)\citenamefont
  {Mukamel}, \citenamefont {Ruffo},\ and\ \citenamefont
  {Schreiber}}]{Schreiber}%
  \BibitemOpen
  \bibfield  {author} {\bibinfo {author} {\bibfnamefont {D.}~\bibnamefont
  {Mukamel}}, \bibinfo {author} {\bibfnamefont {S.}~\bibnamefont {Ruffo}},\
  and\ \bibinfo {author} {\bibfnamefont {N.}~\bibnamefont {Schreiber}},\
  }\bibfield  {title} {\bibinfo {title} {Breaking of ergodicity and long
  relaxation times in systems with long-range interactions},\ }\href
  {https://doi.org/10.1103/PhysRevLett.95.240604} {\bibfield  {journal}
  {\bibinfo  {journal} {Phys. Rev. Lett.}\ }\textbf {\bibinfo {volume} {95}},\
  \bibinfo {pages} {240604} (\bibinfo {year} {2005})}\BibitemShut {NoStop}%
\bibitem [{\citenamefont {Borgonovi}\ \emph {et~al.}(2004)\citenamefont
  {Borgonovi}, \citenamefont {Celardo}, \citenamefont {Maianti},\ and\
  \citenamefont {Pedersoli}}]{Borgonovi}%
  \BibitemOpen
  \bibfield  {author} {\bibinfo {author} {\bibfnamefont {F.}~\bibnamefont
  {Borgonovi}}, \bibinfo {author} {\bibfnamefont {G.~L.}\ \bibnamefont
  {Celardo}}, \bibinfo {author} {\bibfnamefont {M.}~\bibnamefont {Maianti}},\
  and\ \bibinfo {author} {\bibfnamefont {E.}~\bibnamefont {Pedersoli}},\
  }\bibfield  {title} {\bibinfo {title} {Broken ergodicity in classically
  chaotic spin systems},\ }\href
  {https://doi.org/10.1023/b:joss.0000041745.62340.00} {\bibfield  {journal}
  {\bibinfo  {journal} {Journal of Statistical Physics}\ }\textbf {\bibinfo
  {volume} {116}},\ \bibinfo {pages} {1435} (\bibinfo {year}
  {2004})}\BibitemShut {NoStop}%
\bibitem [{\citenamefont
  {Kastner}(2010{\natexlab{a}})}]{kastner2010nonequivalence}%
  \BibitemOpen
  \bibfield  {author} {\bibinfo {author} {\bibfnamefont {M.}~\bibnamefont
  {Kastner}},\ }\bibfield  {title} {\bibinfo {title} {{Nonequivalence of
  Ensembles for Long-Range Quantum Spin Systems in Optical Lattices}},\
  }\href@noop {} {\bibfield  {journal} {\bibinfo  {journal} {Phys. Rev. Lett.}\
  }\textbf {\bibinfo {volume} {104}},\ \bibinfo {pages} {240403} (\bibinfo
  {year} {2010}{\natexlab{a}})}\BibitemShut {NoStop}%
\bibitem [{\citenamefont
  {Kastner}(2010{\natexlab{b}})}]{kastner2010nonequivalence2}%
  \BibitemOpen
  \bibfield  {author} {\bibinfo {author} {\bibfnamefont {M.}~\bibnamefont
  {Kastner}},\ }\bibfield  {title} {\bibinfo {title} {Nonequivalence of
  ensembles in the curie{\textendash}weiss anisotropic quantum heisenberg
  model},\ }\href {https://doi.org/10.1088/1742-5468/2010/07/p07006} {\bibfield
   {journal} {\bibinfo  {journal} {J. Stat. Phys.}\ }\textbf {\bibinfo {volume}
  {2010}},\ \bibinfo {pages} {P07006} (\bibinfo {year}
  {2010}{\natexlab{b}})}\BibitemShut {NoStop}%
\bibitem [{\citenamefont {Russomanno}\ \emph {et~al.}(2021)\citenamefont
  {Russomanno}, \citenamefont {Fava},\ and\ \citenamefont
  {Heyl}}]{russomanno2021chaos}%
  \BibitemOpen
  \bibfield  {author} {\bibinfo {author} {\bibfnamefont {A.}~\bibnamefont
  {Russomanno}}, \bibinfo {author} {\bibfnamefont {M.}~\bibnamefont {Fava}},\
  and\ \bibinfo {author} {\bibfnamefont {M.}~\bibnamefont {Heyl}},\ }\bibfield
  {title} {\bibinfo {title} {Quantum chaos and ensemble inequivalence of
  quantum long-range ising chains},\ }\href
  {https://doi.org/10.1103/PhysRevB.104.094309} {\bibfield  {journal} {\bibinfo
   {journal} {Phys. Rev. B}\ }\textbf {\bibinfo {volume} {104}},\ \bibinfo
  {pages} {094309} (\bibinfo {year} {2021})}\BibitemShut {NoStop}%
\bibitem [{\citenamefont {Del~Re}\ \emph {et~al.}(2016)\citenamefont {Del~Re},
  \citenamefont {Fabrizio},\ and\ \citenamefont {Tosatti}}]{Delre}%
  \BibitemOpen
  \bibfield  {author} {\bibinfo {author} {\bibfnamefont {L.}~\bibnamefont
  {Del~Re}}, \bibinfo {author} {\bibfnamefont {M.}~\bibnamefont {Fabrizio}},\
  and\ \bibinfo {author} {\bibfnamefont {E.}~\bibnamefont {Tosatti}},\
  }\bibfield  {title} {\bibinfo {title} {Nonequilibrium and nonhomogeneous
  phenomena around a first-order quantum phase transition},\ }\href
  {https://doi.org/10.1103/PhysRevB.93.125131} {\bibfield  {journal} {\bibinfo
  {journal} {Phys. Rev. B}\ }\textbf {\bibinfo {volume} {93}},\ \bibinfo
  {pages} {125131} (\bibinfo {year} {2016})}\BibitemShut {NoStop}%
\bibitem [{\citenamefont {Petrov}(2014)}]{petrov2014elastic}%
  \BibitemOpen
  \bibfield  {author} {\bibinfo {author} {\bibfnamefont {D.~S.}\ \bibnamefont
  {Petrov}},\ }\bibfield  {title} {\bibinfo {title} {Elastic multibody
  interactions on a lattice},\ }\href
  {https://doi.org/10.1103/PhysRevA.90.021601} {\bibfield  {journal} {\bibinfo
  {journal} {Phys. Rev. A}\ }\textbf {\bibinfo {volume} {90}},\ \bibinfo
  {pages} {021601} (\bibinfo {year} {2014})}\BibitemShut {NoStop}%
\bibitem [{\citenamefont {{Goban}}\ \emph {et~al.}(2018)\citenamefont
  {{Goban}}, \citenamefont {{Hutson}}, \citenamefont {{Marti}}, \citenamefont
  {{Campbell}}, \citenamefont {{Perlin}}, \citenamefont {{Julienne}},
  \citenamefont {{D'Incao}}, \citenamefont {{Rey}},\ and\ \citenamefont
  {{Ye}}}]{goban2018emergence}%
  \BibitemOpen
  \bibfield  {author} {\bibinfo {author} {\bibfnamefont {A.}~\bibnamefont
  {{Goban}}}, \bibinfo {author} {\bibfnamefont {R.~B.}\ \bibnamefont
  {{Hutson}}}, \bibinfo {author} {\bibfnamefont {G.~E.}\ \bibnamefont
  {{Marti}}}, \bibinfo {author} {\bibfnamefont {S.~L.}\ \bibnamefont
  {{Campbell}}}, \bibinfo {author} {\bibfnamefont {M.~A.}\ \bibnamefont
  {{Perlin}}}, \bibinfo {author} {\bibfnamefont {P.~S.}\ \bibnamefont
  {{Julienne}}}, \bibinfo {author} {\bibfnamefont {J.~P.}\ \bibnamefont
  {{D'Incao}}}, \bibinfo {author} {\bibfnamefont {A.~M.}\ \bibnamefont
  {{Rey}}},\ and\ \bibinfo {author} {\bibfnamefont {J.}~\bibnamefont {{Ye}}},\
  }\bibfield  {title} {\bibinfo {title} {{Emergence of multi-body interactions
  in a fermionic lattice clock}},\ }\href
  {https://doi.org/10.1038/s41586-018-0661-6} {\bibfield  {journal} {\bibinfo
  {journal} {\nat}\ }\textbf {\bibinfo {volume} {563}},\ \bibinfo {pages} {369}
  (\bibinfo {year} {2018})}\BibitemShut {NoStop}%
\bibitem [{\citenamefont {Zwerger}(2019)}]{zwerger2019quantum}%
  \BibitemOpen
  \bibfield  {author} {\bibinfo {author} {\bibfnamefont {W.}~\bibnamefont
  {Zwerger}},\ }\bibfield  {title} {\bibinfo {title} {Quantum-unbinding near a
  zero temperature liquid–gas transition},\ }\href
  {https://doi.org/10.1088/1742-5468/ab3ccc} {\bibfield  {journal} {\bibinfo
  {journal} {Journal of Statistical Mechanics: Theory and Experiment}\ }\textbf
  {\bibinfo {volume} {2019}},\ \bibinfo {pages} {103104} (\bibinfo {year}
  {2019})}\BibitemShut {NoStop}%
\bibitem [{\citenamefont {Griesmaier}\ \emph {et~al.}(2005)\citenamefont
  {Griesmaier}, \citenamefont {Werner}, \citenamefont {Hensler}, \citenamefont
  {Stuhler},\ and\ \citenamefont {Pfau}}]{griesmaier2005bose}%
  \BibitemOpen
  \bibfield  {author} {\bibinfo {author} {\bibfnamefont {A.}~\bibnamefont
  {Griesmaier}}, \bibinfo {author} {\bibfnamefont {J.}~\bibnamefont {Werner}},
  \bibinfo {author} {\bibfnamefont {S.}~\bibnamefont {Hensler}}, \bibinfo
  {author} {\bibfnamefont {J.}~\bibnamefont {Stuhler}},\ and\ \bibinfo {author}
  {\bibfnamefont {T.}~\bibnamefont {Pfau}},\ }\bibfield  {title} {\bibinfo
  {title} {Bose-einstein condensation of chromium},\ }\href
  {https://doi.org/10.1103/PhysRevLett.94.160401} {\bibfield  {journal}
  {\bibinfo  {journal} {Phys. Rev. Lett.}\ }\textbf {\bibinfo {volume} {94}},\
  \bibinfo {pages} {160401} (\bibinfo {year} {2005})}\BibitemShut {NoStop}%
\bibitem [{\citenamefont {{Micheli}}\ \emph {et~al.}(2006)\citenamefont
  {{Micheli}}, \citenamefont {{Brennen}},\ and\ \citenamefont
  {{Zoller}}}]{micheli2006toolbox}%
  \BibitemOpen
  \bibfield  {author} {\bibinfo {author} {\bibfnamefont {A.}~\bibnamefont
  {{Micheli}}}, \bibinfo {author} {\bibfnamefont {G.~K.}\ \bibnamefont
  {{Brennen}}},\ and\ \bibinfo {author} {\bibfnamefont {P.}~\bibnamefont
  {{Zoller}}},\ }\bibfield  {title} {\bibinfo {title} {{A toolbox for
  lattice-spin models with polar molecules}},\ }\href
  {https://doi.org/10.1038/nphys287} {\bibfield  {journal} {\bibinfo  {journal}
  {Nat. Phys.}\ }\textbf {\bibinfo {volume} {2}},\ \bibinfo {pages} {341}
  (\bibinfo {year} {2006})}\BibitemShut {NoStop}%
\bibitem [{\citenamefont {Ni}\ \emph {et~al.}(2008)\citenamefont {Ni},
  \citenamefont {Ospelkaus}, \citenamefont {de~Miranda}, \citenamefont {Pe'er},
  \citenamefont {Neyenhuis}, \citenamefont {Zirbel}, \citenamefont
  {Kotochigova}, \citenamefont {Julienne}, \citenamefont {Jin},\ and\
  \citenamefont {Ye}}]{ni2008high}%
  \BibitemOpen
  \bibfield  {author} {\bibinfo {author} {\bibfnamefont {K.-K.}\ \bibnamefont
  {Ni}}, \bibinfo {author} {\bibfnamefont {S.}~\bibnamefont {Ospelkaus}},
  \bibinfo {author} {\bibfnamefont {M.~H.~G.}\ \bibnamefont {de~Miranda}},
  \bibinfo {author} {\bibfnamefont {A.}~\bibnamefont {Pe'er}}, \bibinfo
  {author} {\bibfnamefont {B.}~\bibnamefont {Neyenhuis}}, \bibinfo {author}
  {\bibfnamefont {J.~J.}\ \bibnamefont {Zirbel}}, \bibinfo {author}
  {\bibfnamefont {S.}~\bibnamefont {Kotochigova}}, \bibinfo {author}
  {\bibfnamefont {P.~S.}\ \bibnamefont {Julienne}}, \bibinfo {author}
  {\bibfnamefont {D.~S.}\ \bibnamefont {Jin}},\ and\ \bibinfo {author}
  {\bibfnamefont {J.}~\bibnamefont {Ye}},\ }\bibfield  {title} {\bibinfo
  {title} {A high phase-space-density gas of polar molecules},\ }\href
  {https://doi.org/10.1126/science.1163861} {\bibfield  {journal} {\bibinfo
  {journal} {Science}\ }\textbf {\bibinfo {volume} {322}},\ \bibinfo {pages}
  {231} (\bibinfo {year} {2008})}\BibitemShut {NoStop}%
\bibitem [{\citenamefont {Morrison}\ and\ \citenamefont
  {Parkins}(2008{\natexlab{a}})}]{morrison2008dynamical}%
  \BibitemOpen
  \bibfield  {author} {\bibinfo {author} {\bibfnamefont {S.}~\bibnamefont
  {Morrison}}\ and\ \bibinfo {author} {\bibfnamefont {A.~S.}\ \bibnamefont
  {Parkins}},\ }\bibfield  {title} {\bibinfo {title} {Dynamical quantum phase
  transitions in the dissipative {L}ipkin-{M}eshkov-{G}lick model with proposed
  realization in optical cavity {QED}},\ }\href
  {https://doi.org/10.1103/PhysRevLett.100.040403} {\bibfield  {journal}
  {\bibinfo  {journal} {Phys. Rev. Lett.}\ }\textbf {\bibinfo {volume} {100}},\
  \bibinfo {pages} {040403} (\bibinfo {year} {2008}{\natexlab{a}})}\BibitemShut
  {NoStop}%
\bibitem [{\citenamefont {Larson}(2010)}]{larson2010circuit}%
  \BibitemOpen
  \bibfield  {author} {\bibinfo {author} {\bibfnamefont {J.}~\bibnamefont
  {Larson}},\ }\bibfield  {title} {\bibinfo {title} {Circuit qed scheme for the
  realization of the lipkin-meshkov-glick model},\ }\href
  {https://doi.org/10.1209/0295-5075/90/54001} {\bibfield  {journal} {\bibinfo
  {journal} {EPL (Europhysics Letters)}\ }\textbf {\bibinfo {volume} {90}},\
  \bibinfo {pages} {54001} (\bibinfo {year} {2010})}\BibitemShut {NoStop}%
\bibitem [{\citenamefont {{Keller}}\ \emph {et~al.}(2018)\citenamefont
  {{Keller}}, \citenamefont {{Torggler}}, \citenamefont {{J{\"a}ger}},
  \citenamefont {{Sch{\"u}tz}}, \citenamefont {{Ritsch}},\ and\ \citenamefont
  {{Morigi}}}]{keller2018quenches}%
  \BibitemOpen
  \bibfield  {author} {\bibinfo {author} {\bibfnamefont {T.}~\bibnamefont
  {{Keller}}}, \bibinfo {author} {\bibfnamefont {V.}~\bibnamefont
  {{Torggler}}}, \bibinfo {author} {\bibfnamefont {S.~B.}\ \bibnamefont
  {{J{\"a}ger}}}, \bibinfo {author} {\bibfnamefont {S.}~\bibnamefont
  {{Sch{\"u}tz}}}, \bibinfo {author} {\bibfnamefont {H.}~\bibnamefont
  {{Ritsch}}},\ and\ \bibinfo {author} {\bibfnamefont {G.}~\bibnamefont
  {{Morigi}}},\ }\bibfield  {title} {\bibinfo {title} {{Quenches across the
  self-organization transition in multimode cavities}},\ }\href
  {https://doi.org/10.1088/1367-2630/aaa161} {\bibfield  {journal} {\bibinfo
  {journal} {New Journal of Physics}\ }\textbf {\bibinfo {volume} {20}},\
  \bibinfo {eid} {025004} (\bibinfo {year} {2018})}\BibitemShut {NoStop}%
\bibitem [{\citenamefont {Wu}\ \emph {et~al.}(2023)\citenamefont {Wu},
  \citenamefont {Fan}, \citenamefont {Zhang}, \citenamefont {Qi},\ and\
  \citenamefont {Wu}}]{wu2023signatures}%
  \BibitemOpen
  \bibfield  {author} {\bibinfo {author} {\bibfnamefont {Z.}~\bibnamefont
  {Wu}}, \bibinfo {author} {\bibfnamefont {J.}~\bibnamefont {Fan}}, \bibinfo
  {author} {\bibfnamefont {X.}~\bibnamefont {Zhang}}, \bibinfo {author}
  {\bibfnamefont {J.}~\bibnamefont {Qi}},\ and\ \bibinfo {author}
  {\bibfnamefont {H.}~\bibnamefont {Wu}},\ }\bibfield  {title} {\bibinfo
  {title} {Signatures of prethermalization in a quenched cavity-mediated
  long-range interacting fermi gas},\ }\href
  {https://doi.org/10.1103/PhysRevLett.131.243401} {\bibfield  {journal}
  {\bibinfo  {journal} {Phys. Rev. Lett.}\ }\textbf {\bibinfo {volume} {131}},\
  \bibinfo {pages} {243401} (\bibinfo {year} {2023})}\BibitemShut {NoStop}%
\bibitem [{\citenamefont {Albash}\ and\ \citenamefont
  {Lidar}(2018)}]{albash2018adiabatic}%
  \BibitemOpen
  \bibfield  {author} {\bibinfo {author} {\bibfnamefont {T.}~\bibnamefont
  {Albash}}\ and\ \bibinfo {author} {\bibfnamefont {D.~A.}\ \bibnamefont
  {Lidar}},\ }\bibfield  {title} {\bibinfo {title} {Adiabatic quantum
  computation},\ }\href {https://doi.org/10.1103/RevModPhys.90.015002}
  {\bibfield  {journal} {\bibinfo  {journal} {Rev. Mod. Phys.}\ }\textbf
  {\bibinfo {volume} {90}},\ \bibinfo {pages} {015002} (\bibinfo {year}
  {2018})}\BibitemShut {NoStop}%
\bibitem [{\citenamefont {Lipkin}\ \emph {et~al.}(1965)\citenamefont {Lipkin},
  \citenamefont {Meshkov},\ and\ \citenamefont {Glick}}]{lipkin1965validity}%
  \BibitemOpen
  \bibfield  {author} {\bibinfo {author} {\bibfnamefont {H.~J.}\ \bibnamefont
  {Lipkin}}, \bibinfo {author} {\bibfnamefont {N.}~\bibnamefont {Meshkov}},\
  and\ \bibinfo {author} {\bibfnamefont {A.~J.}\ \bibnamefont {Glick}},\
  }\bibfield  {title} {\bibinfo {title} {{Validity of many-body approximation
  methods for a solvable model}},\ }\href@noop {} {\bibfield  {journal}
  {\bibinfo  {journal} {Nuclear Physics}\ }\textbf {\bibinfo {volume} {62}},\
  \bibinfo {pages} {188} (\bibinfo {year} {1965})}\BibitemShut {NoStop}%
\bibitem [{\citenamefont {{Meshkov}}\ \emph {et~al.}(1965)\citenamefont
  {{Meshkov}}, \citenamefont {{Glick}},\ and\ \citenamefont
  {{Lipkin}}}]{meshkov1965validity}%
  \BibitemOpen
  \bibfield  {author} {\bibinfo {author} {\bibfnamefont {N.}~\bibnamefont
  {{Meshkov}}}, \bibinfo {author} {\bibfnamefont {A.~J.}\ \bibnamefont
  {{Glick}}},\ and\ \bibinfo {author} {\bibfnamefont {H.~J.}\ \bibnamefont
  {{Lipkin}}},\ }\bibfield  {title} {\bibinfo {title} {{Validity of many-body
  approximation methods for a solvable model. (II). Linearization
  procedures}},\ }\href {https://doi.org/10.1016/0029-5582(65)90863-1}
  {\bibfield  {journal} {\bibinfo  {journal} {Nuclear Physics}\ }\textbf
  {\bibinfo {volume} {62}},\ \bibinfo {pages} {199} (\bibinfo {year}
  {1965})}\BibitemShut {NoStop}%
\bibitem [{\citenamefont {{Glick}}\ \emph {et~al.}(1965)\citenamefont
  {{Glick}}, \citenamefont {{Lipkin}},\ and\ \citenamefont
  {{Meshkov}}}]{glick1965validity}%
  \BibitemOpen
  \bibfield  {author} {\bibinfo {author} {\bibfnamefont {A.~J.}\ \bibnamefont
  {{Glick}}}, \bibinfo {author} {\bibfnamefont {H.~J.}\ \bibnamefont
  {{Lipkin}}},\ and\ \bibinfo {author} {\bibfnamefont {N.}~\bibnamefont
  {{Meshkov}}},\ }\bibfield  {title} {\bibinfo {title} {{Validity of many-body
  approximation methods for a solvable model. (III). Diagram summations}},\
  }\href {https://doi.org/10.1016/0029-5582(65)90864-3} {\bibfield  {journal}
  {\bibinfo  {journal} {Nuclear Physics}\ }\textbf {\bibinfo {volume} {62}},\
  \bibinfo {pages} {211} (\bibinfo {year} {1965})}\BibitemShut {NoStop}%
\bibitem [{\citenamefont {Dicke}(1954)}]{dicke1954coherence}%
  \BibitemOpen
  \bibfield  {author} {\bibinfo {author} {\bibfnamefont {R.~H.}\ \bibnamefont
  {Dicke}},\ }\bibfield  {title} {\bibinfo {title} {Coherence in spontaneous
  radiation processes},\ }\href {https://doi.org/10.1103/PhysRev.93.99}
  {\bibfield  {journal} {\bibinfo  {journal} {Phys. Rev.}\ }\textbf {\bibinfo
  {volume} {93}},\ \bibinfo {pages} {99} (\bibinfo {year} {1954})}\BibitemShut
  {NoStop}%
\bibitem [{\citenamefont {Baumann}\ \emph {et~al.}(2010)\citenamefont
  {Baumann}, \citenamefont {Guerlin}, \citenamefont {Brennecke},\ and\
  \citenamefont {Esslinger}}]{baumann2010dicke}%
  \BibitemOpen
  \bibfield  {author} {\bibinfo {author} {\bibfnamefont {K.}~\bibnamefont
  {Baumann}}, \bibinfo {author} {\bibfnamefont {C.}~\bibnamefont {Guerlin}},
  \bibinfo {author} {\bibfnamefont {F.}~\bibnamefont {Brennecke}},\ and\
  \bibinfo {author} {\bibfnamefont {T.}~\bibnamefont {Esslinger}},\ }\bibfield
  {title} {\bibinfo {title} {{Dicke quantum phase transition with a superfluid
  gas in an optical cavity}},\ }\href@noop {} {\bibfield  {journal} {\bibinfo
  {journal} {Nature}\ }\textbf {\bibinfo {volume} {464}},\ \bibinfo {pages}
  {1301} (\bibinfo {year} {2010})}\BibitemShut {NoStop}%
\bibitem [{\citenamefont {Landig}\ \emph {et~al.}(2015)\citenamefont {Landig},
  \citenamefont {Brennecke}, \citenamefont {Mottl}, \citenamefont {Donner},\
  and\ \citenamefont {Esslinger}}]{landig2015measuring}%
  \BibitemOpen
  \bibfield  {author} {\bibinfo {author} {\bibfnamefont {R.}~\bibnamefont
  {Landig}}, \bibinfo {author} {\bibfnamefont {F.}~\bibnamefont {Brennecke}},
  \bibinfo {author} {\bibfnamefont {R.}~\bibnamefont {Mottl}}, \bibinfo
  {author} {\bibfnamefont {T.}~\bibnamefont {Donner}},\ and\ \bibinfo {author}
  {\bibfnamefont {T.}~\bibnamefont {Esslinger}},\ }\bibfield  {title} {\bibinfo
  {title} {{Measuring the dynamic structure factor of a quantum gas undergoing
  a structural phase transition}},\ }\href@noop {} {\bibfield  {journal}
  {\bibinfo  {journal} {Nat. Comm.}\ }\textbf {\bibinfo {volume} {6}},\
  \bibinfo {pages} {7046} (\bibinfo {year} {2015})}\BibitemShut {NoStop}%
\bibitem [{\citenamefont {Reslen}\ \emph {et~al.}(2005)\citenamefont {Reslen},
  \citenamefont {Quiroga},\ and\ \citenamefont {Johnson}}]{reslen2005direct}%
  \BibitemOpen
  \bibfield  {author} {\bibinfo {author} {\bibfnamefont {J.}~\bibnamefont
  {Reslen}}, \bibinfo {author} {\bibfnamefont {L.}~\bibnamefont {Quiroga}},\
  and\ \bibinfo {author} {\bibfnamefont {N.~F.}\ \bibnamefont {Johnson}},\
  }\bibfield  {title} {\bibinfo {title} {Direct equivalence between quantum
  phase transition phenomena in radiation-matter and magnetic systems: Scaling
  of entanglement},\ }\href {https://doi.org/10.1209/epl/i2004-10313-4}
  {\bibfield  {journal} {\bibinfo  {journal} {EPL}\ }\textbf {\bibinfo {volume}
  {69}},\ \bibinfo {pages} {8} (\bibinfo {year} {2005})}\BibitemShut {NoStop}%
\bibitem [{\citenamefont {Sch{\"u}tz}\ \emph {et~al.}(2015)\citenamefont
  {Sch{\"u}tz}, \citenamefont {J{\"a}ger},\ and\ \citenamefont
  {Morigi}}]{schutz2015thermodynamics}%
  \BibitemOpen
  \bibfield  {author} {\bibinfo {author} {\bibfnamefont {S.}~\bibnamefont
  {Sch{\"u}tz}}, \bibinfo {author} {\bibfnamefont {S.~B.}\ \bibnamefont
  {J{\"a}ger}},\ and\ \bibinfo {author} {\bibfnamefont {G.}~\bibnamefont
  {Morigi}},\ }\bibfield  {title} {\bibinfo {title} {{Thermodynamics and
  dynamics of atomic self-organization in an optical cavity}},\ }\href@noop {}
  {\bibfield  {journal} {\bibinfo  {journal} {Phys. Rev. A}\ }\textbf {\bibinfo
  {volume} {92}},\ \bibinfo {pages} {063808} (\bibinfo {year}
  {2015})}\BibitemShut {NoStop}%
\bibitem [{\citenamefont {Leroux}\ \emph {et~al.}(2010)\citenamefont {Leroux},
  \citenamefont {Schleier-Smith},\ and\ \citenamefont
  {Vuleti\ifmmode~\acute{c}\else \'{c}\fi{}}}]{leroux2010implementation}%
  \BibitemOpen
  \bibfield  {author} {\bibinfo {author} {\bibfnamefont {I.~D.}\ \bibnamefont
  {Leroux}}, \bibinfo {author} {\bibfnamefont {M.~H.}\ \bibnamefont
  {Schleier-Smith}},\ and\ \bibinfo {author} {\bibfnamefont {V.}~\bibnamefont
  {Vuleti\ifmmode~\acute{c}\else \'{c}\fi{}}},\ }\bibfield  {title} {\bibinfo
  {title} {Implementation of cavity squeezing of a collective atomic spin},\
  }\href {https://doi.org/10.1103/PhysRevLett.104.073602} {\bibfield  {journal}
  {\bibinfo  {journal} {Phys. Rev. Lett.}\ }\textbf {\bibinfo {volume} {104}},\
  \bibinfo {pages} {073602} (\bibinfo {year} {2010})}\BibitemShut {NoStop}%
\bibitem [{\citenamefont {Bentsen}\ \emph {et~al.}(2019)\citenamefont
  {Bentsen}, \citenamefont {Potirniche}, \citenamefont {Bulchandani},
  \citenamefont {Scaffidi}, \citenamefont {Cao}, \citenamefont {Qi},
  \citenamefont {Schleier-Smith},\ and\ \citenamefont
  {Altman}}]{bentsen2019integrable}%
  \BibitemOpen
  \bibfield  {author} {\bibinfo {author} {\bibfnamefont {G.}~\bibnamefont
  {Bentsen}}, \bibinfo {author} {\bibfnamefont {I.-D.}\ \bibnamefont
  {Potirniche}}, \bibinfo {author} {\bibfnamefont {V.~B.}\ \bibnamefont
  {Bulchandani}}, \bibinfo {author} {\bibfnamefont {T.}~\bibnamefont
  {Scaffidi}}, \bibinfo {author} {\bibfnamefont {X.}~\bibnamefont {Cao}},
  \bibinfo {author} {\bibfnamefont {X.-L.}\ \bibnamefont {Qi}}, \bibinfo
  {author} {\bibfnamefont {M.}~\bibnamefont {Schleier-Smith}},\ and\ \bibinfo
  {author} {\bibfnamefont {E.}~\bibnamefont {Altman}},\ }\bibfield  {title}
  {\bibinfo {title} {Integrable and chaotic dynamics of spins coupled to an
  optical cavity},\ }\href {https://doi.org/10.1103/PhysRevX.9.041011}
  {\bibfield  {journal} {\bibinfo  {journal} {Phys. Rev. X}\ }\textbf {\bibinfo
  {volume} {9}},\ \bibinfo {pages} {041011} (\bibinfo {year}
  {2019})}\BibitemShut {NoStop}%
\bibitem [{\citenamefont {Davis}\ \emph {et~al.}(2019)\citenamefont {Davis},
  \citenamefont {Bentsen}, \citenamefont {Homeier}, \citenamefont {Li},\ and\
  \citenamefont {Schleier-Smith}}]{davis2019photon}%
  \BibitemOpen
  \bibfield  {author} {\bibinfo {author} {\bibfnamefont {E.~J.}\ \bibnamefont
  {Davis}}, \bibinfo {author} {\bibfnamefont {G.}~\bibnamefont {Bentsen}},
  \bibinfo {author} {\bibfnamefont {L.}~\bibnamefont {Homeier}}, \bibinfo
  {author} {\bibfnamefont {T.}~\bibnamefont {Li}},\ and\ \bibinfo {author}
  {\bibfnamefont {M.~H.}\ \bibnamefont {Schleier-Smith}},\ }\bibfield  {title}
  {\bibinfo {title} {Photon-mediated spin-exchange dynamics of spin-1 atoms},\
  }\href {https://doi.org/10.1103/PhysRevLett.122.010405} {\bibfield  {journal}
  {\bibinfo  {journal} {Phys. Rev. Lett.}\ }\textbf {\bibinfo {volume} {122}},\
  \bibinfo {pages} {010405} (\bibinfo {year} {2019})}\BibitemShut {NoStop}%
\bibitem [{\citenamefont {Davis}\ \emph {et~al.}(2020)\citenamefont {Davis},
  \citenamefont {Periwal}, \citenamefont {Cooper}, \citenamefont {Bentsen},
  \citenamefont {Evered}, \citenamefont {Van~Kirk},\ and\ \citenamefont
  {Schleier-Smith}}]{davis2020protecting}%
  \BibitemOpen
  \bibfield  {author} {\bibinfo {author} {\bibfnamefont {E.~J.}\ \bibnamefont
  {Davis}}, \bibinfo {author} {\bibfnamefont {A.}~\bibnamefont {Periwal}},
  \bibinfo {author} {\bibfnamefont {E.~S.}\ \bibnamefont {Cooper}}, \bibinfo
  {author} {\bibfnamefont {G.}~\bibnamefont {Bentsen}}, \bibinfo {author}
  {\bibfnamefont {S.~J.}\ \bibnamefont {Evered}}, \bibinfo {author}
  {\bibfnamefont {K.}~\bibnamefont {Van~Kirk}},\ and\ \bibinfo {author}
  {\bibfnamefont {M.~H.}\ \bibnamefont {Schleier-Smith}},\ }\bibfield  {title}
  {\bibinfo {title} {Protecting spin coherence in a tunable heisenberg model},\
  }\href {https://doi.org/10.1103/PhysRevLett.125.060402} {\bibfield  {journal}
  {\bibinfo  {journal} {Phys. Rev. Lett.}\ }\textbf {\bibinfo {volume} {125}},\
  \bibinfo {pages} {060402} (\bibinfo {year} {2020})}\BibitemShut {NoStop}%
\bibitem [{\citenamefont {Gallem\'{\i}}\ \emph {et~al.}(2016)\citenamefont
  {Gallem\'{\i}}, \citenamefont {Queralt\'o}, \citenamefont {Guilleumas},
  \citenamefont {Mayol},\ and\ \citenamefont {Sanpera}}]{gallemi2016quantum}%
  \BibitemOpen
  \bibfield  {author} {\bibinfo {author} {\bibfnamefont {A.}~\bibnamefont
  {Gallem\'{\i}}}, \bibinfo {author} {\bibfnamefont {G.}~\bibnamefont
  {Queralt\'o}}, \bibinfo {author} {\bibfnamefont {M.}~\bibnamefont
  {Guilleumas}}, \bibinfo {author} {\bibfnamefont {R.}~\bibnamefont {Mayol}},\
  and\ \bibinfo {author} {\bibfnamefont {A.}~\bibnamefont {Sanpera}},\
  }\bibfield  {title} {\bibinfo {title} {Quantum spin models with mesoscopic
  bose-einstein condensates},\ }\href
  {https://doi.org/10.1103/PhysRevA.94.063626} {\bibfield  {journal} {\bibinfo
  {journal} {Phys. Rev. A}\ }\textbf {\bibinfo {volume} {94}},\ \bibinfo
  {pages} {063626} (\bibinfo {year} {2016})}\BibitemShut {NoStop}%
\bibitem [{\citenamefont {Ho}(1998)}]{ho1998spinor}%
  \BibitemOpen
  \bibfield  {author} {\bibinfo {author} {\bibfnamefont {T.-L.}\ \bibnamefont
  {Ho}},\ }\bibfield  {title} {\bibinfo {title} {Spinor bose condensates in
  optical traps},\ }\href {https://doi.org/10.1103/PhysRevLett.81.742}
  {\bibfield  {journal} {\bibinfo  {journal} {Phys. Rev. Lett.}\ }\textbf
  {\bibinfo {volume} {81}},\ \bibinfo {pages} {742} (\bibinfo {year}
  {1998})}\BibitemShut {NoStop}%
\bibitem [{\citenamefont {Ohmi}\ and\ \citenamefont
  {Machida}(1998)}]{ohmi1998bose}%
  \BibitemOpen
  \bibfield  {author} {\bibinfo {author} {\bibfnamefont {T.}~\bibnamefont
  {Ohmi}}\ and\ \bibinfo {author} {\bibfnamefont {K.}~\bibnamefont {Machida}},\
  }\bibfield  {title} {\bibinfo {title} {Bose-einstein condensation with
  internal degrees of freedom in alkali atom gases},\ }\href
  {https://doi.org/10.1143/JPSJ.67.1822} {\bibfield  {journal} {\bibinfo
  {journal} {Journal of the Physical Society of Japan}\ }\textbf {\bibinfo
  {volume} {67}},\ \bibinfo {pages} {1822} (\bibinfo {year} {1998})},\ \Eprint
  {https://arxiv.org/abs/https://doi.org/10.1143/JPSJ.67.1822}
  {https://doi.org/10.1143/JPSJ.67.1822} \BibitemShut {NoStop}%
\bibitem [{\citenamefont {{Stenger}}\ \emph {et~al.}(1998)\citenamefont
  {{Stenger}}, \citenamefont {{Inouye}}, \citenamefont {{Stamper-Kurn}},
  \citenamefont {{Miesner}}, \citenamefont {{Chikkatur}},\ and\ \citenamefont
  {{Ketterle}}}]{stenger1998spin}%
  \BibitemOpen
  \bibfield  {author} {\bibinfo {author} {\bibfnamefont {J.}~\bibnamefont
  {{Stenger}}}, \bibinfo {author} {\bibfnamefont {S.}~\bibnamefont {{Inouye}}},
  \bibinfo {author} {\bibfnamefont {D.~M.}\ \bibnamefont {{Stamper-Kurn}}},
  \bibinfo {author} {\bibfnamefont {H.~J.}\ \bibnamefont {{Miesner}}}, \bibinfo
  {author} {\bibfnamefont {A.~P.}\ \bibnamefont {{Chikkatur}}},\ and\ \bibinfo
  {author} {\bibfnamefont {W.}~\bibnamefont {{Ketterle}}},\ }\bibfield  {title}
  {\bibinfo {title} {{Spin domains in ground-state Bose-Einstein
  condensates}},\ }\href {https://doi.org/10.1038/24567} {\bibfield  {journal}
  {\bibinfo  {journal} {\nat}\ }\textbf {\bibinfo {volume} {396}},\ \bibinfo
  {pages} {345} (\bibinfo {year} {1998})}\BibitemShut {NoStop}%
\bibitem [{\citenamefont {Chang}\ \emph {et~al.}(2004)\citenamefont {Chang},
  \citenamefont {Hamley}, \citenamefont {Barrett}, \citenamefont {Sauer},
  \citenamefont {Fortier}, \citenamefont {Zhang}, \citenamefont {You},\ and\
  \citenamefont {Chapman}}]{chang2004observation}%
  \BibitemOpen
  \bibfield  {author} {\bibinfo {author} {\bibfnamefont {M.-S.}\ \bibnamefont
  {Chang}}, \bibinfo {author} {\bibfnamefont {C.~D.}\ \bibnamefont {Hamley}},
  \bibinfo {author} {\bibfnamefont {M.~D.}\ \bibnamefont {Barrett}}, \bibinfo
  {author} {\bibfnamefont {J.~A.}\ \bibnamefont {Sauer}}, \bibinfo {author}
  {\bibfnamefont {K.~M.}\ \bibnamefont {Fortier}}, \bibinfo {author}
  {\bibfnamefont {W.}~\bibnamefont {Zhang}}, \bibinfo {author} {\bibfnamefont
  {L.}~\bibnamefont {You}},\ and\ \bibinfo {author} {\bibfnamefont {M.~S.}\
  \bibnamefont {Chapman}},\ }\bibfield  {title} {\bibinfo {title} {Observation
  of spinor dynamics in optically trapped $^{87}\mathrm{Rb}$ bose-einstein
  condensates},\ }\href {https://doi.org/10.1103/PhysRevLett.92.140403}
  {\bibfield  {journal} {\bibinfo  {journal} {Phys. Rev. Lett.}\ }\textbf
  {\bibinfo {volume} {92}},\ \bibinfo {pages} {140403} (\bibinfo {year}
  {2004})}\BibitemShut {NoStop}%
\bibitem [{\citenamefont {Schmaljohann}\ \emph {et~al.}(2004)\citenamefont
  {Schmaljohann}, \citenamefont {Erhard}, \citenamefont {Kronj\"ager},
  \citenamefont {Kottke}, \citenamefont {van Staa}, \citenamefont
  {Cacciapuoti}, \citenamefont {Arlt}, \citenamefont {Bongs},\ and\
  \citenamefont {Sengstock}}]{schmaljohann2004dynamics}%
  \BibitemOpen
  \bibfield  {author} {\bibinfo {author} {\bibfnamefont {H.}~\bibnamefont
  {Schmaljohann}}, \bibinfo {author} {\bibfnamefont {M.}~\bibnamefont
  {Erhard}}, \bibinfo {author} {\bibfnamefont {J.}~\bibnamefont {Kronj\"ager}},
  \bibinfo {author} {\bibfnamefont {M.}~\bibnamefont {Kottke}}, \bibinfo
  {author} {\bibfnamefont {S.}~\bibnamefont {van Staa}}, \bibinfo {author}
  {\bibfnamefont {L.}~\bibnamefont {Cacciapuoti}}, \bibinfo {author}
  {\bibfnamefont {J.~J.}\ \bibnamefont {Arlt}}, \bibinfo {author}
  {\bibfnamefont {K.}~\bibnamefont {Bongs}},\ and\ \bibinfo {author}
  {\bibfnamefont {K.}~\bibnamefont {Sengstock}},\ }\bibfield  {title} {\bibinfo
  {title} {Dynamics of $f=2$ spinor bose-einstein condensates},\ }\href
  {https://doi.org/10.1103/PhysRevLett.92.040402} {\bibfield  {journal}
  {\bibinfo  {journal} {Phys. Rev. Lett.}\ }\textbf {\bibinfo {volume} {92}},\
  \bibinfo {pages} {040402} (\bibinfo {year} {2004})}\BibitemShut {NoStop}%
\bibitem [{\citenamefont {{Hoang}}\ \emph {et~al.}(2016)\citenamefont
  {{Hoang}}, \citenamefont {{Anquez}}, \citenamefont {{Robbins}}, \citenamefont
  {{Yang}}, \citenamefont {{Land}}, \citenamefont {{Hamley}},\ and\
  \citenamefont {{Chapman}}}]{hoang2016parametric}%
  \BibitemOpen
  \bibfield  {author} {\bibinfo {author} {\bibfnamefont {T.~M.}\ \bibnamefont
  {{Hoang}}}, \bibinfo {author} {\bibfnamefont {M.}~\bibnamefont {{Anquez}}},
  \bibinfo {author} {\bibfnamefont {B.~A.}\ \bibnamefont {{Robbins}}}, \bibinfo
  {author} {\bibfnamefont {X.~Y.}\ \bibnamefont {{Yang}}}, \bibinfo {author}
  {\bibfnamefont {B.~J.}\ \bibnamefont {{Land}}}, \bibinfo {author}
  {\bibfnamefont {C.~D.}\ \bibnamefont {{Hamley}}},\ and\ \bibinfo {author}
  {\bibfnamefont {M.~S.}\ \bibnamefont {{Chapman}}},\ }\bibfield  {title}
  {\bibinfo {title} {{Parametric excitation and squeezing in a many-body spinor
  condensate}},\ }\href {https://doi.org/10.1038/ncomms11233} {\bibfield
  {journal} {\bibinfo  {journal} {Nature Communications}\ }\textbf {\bibinfo
  {volume} {7}},\ \bibinfo {eid} {11233} (\bibinfo {year} {2016})}\BibitemShut
  {NoStop}%
\bibitem [{\citenamefont {Weimer}\ \emph {et~al.}(2010)\citenamefont {Weimer},
  \citenamefont {M{\"u}ller}, \citenamefont {Lesanovsky}, \citenamefont
  {Zoller},\ and\ \citenamefont {B{\"u}chler}}]{weimer2010rydberg}%
  \BibitemOpen
  \bibfield  {author} {\bibinfo {author} {\bibfnamefont {H.}~\bibnamefont
  {Weimer}}, \bibinfo {author} {\bibfnamefont {M.}~\bibnamefont {M{\"u}ller}},
  \bibinfo {author} {\bibfnamefont {I.}~\bibnamefont {Lesanovsky}}, \bibinfo
  {author} {\bibfnamefont {P.}~\bibnamefont {Zoller}},\ and\ \bibinfo {author}
  {\bibfnamefont {H.~P.}\ \bibnamefont {B{\"u}chler}},\ }\bibfield  {title}
  {\bibinfo {title} {A rydberg quantum simulator},\ }\href
  {https://doi.org/10.1038/nphys1614} {\bibfield  {journal} {\bibinfo
  {journal} {Nature Physics}\ }\textbf {\bibinfo {volume} {6}},\ \bibinfo
  {pages} {382} (\bibinfo {year} {2010})}\BibitemShut {NoStop}%
\bibitem [{\citenamefont {Henkel}\ \emph {et~al.}(2010)\citenamefont {Henkel},
  \citenamefont {Nath},\ and\ \citenamefont {Pohl}}]{henkel2010three}%
  \BibitemOpen
  \bibfield  {author} {\bibinfo {author} {\bibfnamefont {N.}~\bibnamefont
  {Henkel}}, \bibinfo {author} {\bibfnamefont {R.}~\bibnamefont {Nath}},\ and\
  \bibinfo {author} {\bibfnamefont {T.}~\bibnamefont {Pohl}},\ }\bibfield
  {title} {\bibinfo {title} {Three-dimensional roton excitations and supersolid
  formation in rydberg-excited bose-einstein condensates},\ }\href
  {https://doi.org/10.1103/PhysRevLett.104.195302} {\bibfield  {journal}
  {\bibinfo  {journal} {Phys. Rev. Lett.}\ }\textbf {\bibinfo {volume} {104}},\
  \bibinfo {pages} {195302} (\bibinfo {year} {2010})}\BibitemShut {NoStop}%
\bibitem [{\citenamefont {Gil}\ \emph {et~al.}(2014)\citenamefont {Gil},
  \citenamefont {Mukherjee}, \citenamefont {Bridge}, \citenamefont {Jones},\
  and\ \citenamefont {Pohl}}]{gil2014spin}%
  \BibitemOpen
  \bibfield  {author} {\bibinfo {author} {\bibfnamefont {L.~I.~R.}\
  \bibnamefont {Gil}}, \bibinfo {author} {\bibfnamefont {R.}~\bibnamefont
  {Mukherjee}}, \bibinfo {author} {\bibfnamefont {E.~M.}\ \bibnamefont
  {Bridge}}, \bibinfo {author} {\bibfnamefont {M.~P.~A.}\ \bibnamefont
  {Jones}},\ and\ \bibinfo {author} {\bibfnamefont {T.}~\bibnamefont {Pohl}},\
  }\bibfield  {title} {\bibinfo {title} {Spin squeezing in a rydberg lattice
  clock},\ }\href {https://doi.org/10.1103/PhysRevLett.112.103601} {\bibfield
  {journal} {\bibinfo  {journal} {Phys. Rev. Lett.}\ }\textbf {\bibinfo
  {volume} {112}},\ \bibinfo {pages} {103601} (\bibinfo {year}
  {2014})}\BibitemShut {NoStop}%
\bibitem [{\citenamefont {Zeiher}\ \emph {et~al.}(2015)\citenamefont {Zeiher},
  \citenamefont {Schau\ss{}}, \citenamefont {Hild}, \citenamefont {Macr\`{\i}},
  \citenamefont {Bloch},\ and\ \citenamefont {Gross}}]{zeiher2015microscopic}%
  \BibitemOpen
  \bibfield  {author} {\bibinfo {author} {\bibfnamefont {J.}~\bibnamefont
  {Zeiher}}, \bibinfo {author} {\bibfnamefont {P.}~\bibnamefont {Schau\ss{}}},
  \bibinfo {author} {\bibfnamefont {S.}~\bibnamefont {Hild}}, \bibinfo {author}
  {\bibfnamefont {T.}~\bibnamefont {Macr\`{\i}}}, \bibinfo {author}
  {\bibfnamefont {I.}~\bibnamefont {Bloch}},\ and\ \bibinfo {author}
  {\bibfnamefont {C.}~\bibnamefont {Gross}},\ }\bibfield  {title} {\bibinfo
  {title} {Microscopic characterization of scalable coherent rydberg
  superatoms},\ }\href {https://doi.org/10.1103/PhysRevX.5.031015} {\bibfield
  {journal} {\bibinfo  {journal} {Phys. Rev. X}\ }\textbf {\bibinfo {volume}
  {5}},\ \bibinfo {pages} {031015} (\bibinfo {year} {2015})}\BibitemShut
  {NoStop}%
\bibitem [{\citenamefont {Jau}\ \emph {et~al.}(2016)\citenamefont {Jau},
  \citenamefont {Hankin}, \citenamefont {Keating}, \citenamefont {Deutsch},\
  and\ \citenamefont {Biedermann}}]{jau2016entangling}%
  \BibitemOpen
  \bibfield  {author} {\bibinfo {author} {\bibfnamefont {Y.~Y.}\ \bibnamefont
  {Jau}}, \bibinfo {author} {\bibfnamefont {A.~M.}\ \bibnamefont {Hankin}},
  \bibinfo {author} {\bibfnamefont {T.}~\bibnamefont {Keating}}, \bibinfo
  {author} {\bibfnamefont {I.~H.}\ \bibnamefont {Deutsch}},\ and\ \bibinfo
  {author} {\bibfnamefont {G.~W.}\ \bibnamefont {Biedermann}},\ }\bibfield
  {title} {\bibinfo {title} {Entangling atomic spins with a rydberg-dressed
  spin-flip blockade},\ }\href {https://doi.org/10.1038/nphys3487} {\bibfield
  {journal} {\bibinfo  {journal} {Nature Physics}\ }\textbf {\bibinfo {volume}
  {12}},\ \bibinfo {pages} {71} (\bibinfo {year} {2016})}\BibitemShut {NoStop}%
\bibitem [{\citenamefont {{Will}}\ \emph {et~al.}(2010)\citenamefont {{Will}},
  \citenamefont {{Best}}, \citenamefont {{Schneider}}, \citenamefont
  {{Hackerm{\"u}ller}}, \citenamefont {{L{\"u}hmann}},\ and\ \citenamefont
  {{Bloch}}}]{will2010time}%
  \BibitemOpen
  \bibfield  {author} {\bibinfo {author} {\bibfnamefont {S.}~\bibnamefont
  {{Will}}}, \bibinfo {author} {\bibfnamefont {T.}~\bibnamefont {{Best}}},
  \bibinfo {author} {\bibfnamefont {U.}~\bibnamefont {{Schneider}}}, \bibinfo
  {author} {\bibfnamefont {L.}~\bibnamefont {{Hackerm{\"u}ller}}}, \bibinfo
  {author} {\bibfnamefont {D.-S.}\ \bibnamefont {{L{\"u}hmann}}},\ and\
  \bibinfo {author} {\bibfnamefont {I.}~\bibnamefont {{Bloch}}},\ }\bibfield
  {title} {\bibinfo {title} {{Time-resolved observation of coherent multi-body
  interactions in quantum phase revivals}},\ }\href
  {https://doi.org/10.1038/nature09036} {\bibfield  {journal} {\bibinfo
  {journal} {Nature}\ }\textbf {\bibinfo {volume} {465}},\ \bibinfo {pages}
  {197} (\bibinfo {year} {2010})}\BibitemShut {NoStop}%
\bibitem [{\citenamefont {{B{\"u}chler}}\ \emph {et~al.}(2007)\citenamefont
  {{B{\"u}chler}}, \citenamefont {{Micheli}},\ and\ \citenamefont
  {{Zoller}}}]{buchler2007three}%
  \BibitemOpen
  \bibfield  {author} {\bibinfo {author} {\bibfnamefont {H.~P.}\ \bibnamefont
  {{B{\"u}chler}}}, \bibinfo {author} {\bibfnamefont {A.}~\bibnamefont
  {{Micheli}}},\ and\ \bibinfo {author} {\bibfnamefont {P.}~\bibnamefont
  {{Zoller}}},\ }\bibfield  {title} {\bibinfo {title} {{Three-body interactions
  with cold polar molecules}},\ }\href {https://doi.org/10.1038/nphys678}
  {\bibfield  {journal} {\bibinfo  {journal} {Nat. Phys.}\ }\textbf {\bibinfo
  {volume} {3}},\ \bibinfo {pages} {726} (\bibinfo {year} {2007})}\BibitemShut
  {NoStop}%
\bibitem [{\citenamefont {Lieb}(1973)}]{Lieb}%
  \BibitemOpen
  \bibfield  {author} {\bibinfo {author} {\bibfnamefont {E.~H.}\ \bibnamefont
  {Lieb}},\ }\bibfield  {title} {\bibinfo {title} {The classical limit of
  quantum spin systems},\ }\href@noop {} {\bibfield  {journal} {\bibinfo
  {journal} {Commun. Math. Phys}\ }\textbf {\bibinfo {volume} {31}},\ \bibinfo
  {pages} {327} (\bibinfo {year} {1973})}\BibitemShut {NoStop}%
\bibitem [{\citenamefont {Granet}(2023)}]{Granet}%
  \BibitemOpen
  \bibfield  {author} {\bibinfo {author} {\bibfnamefont {E.}~\bibnamefont
  {Granet}},\ }\bibfield  {title} {\bibinfo {title} {{Exact mean-field solution
  of a spin chain with short-range and long-range interactions}},\ }\href
  {https://doi.org/10.21468/SciPostPhys.14.5.133} {\bibfield  {journal}
  {\bibinfo  {journal} {SciPost Phys.}\ }\textbf {\bibinfo {volume} {14}},\
  \bibinfo {pages} {133} (\bibinfo {year} {2023})}\BibitemShut {NoStop}%
\bibitem [{\citenamefont {Morrison}\ and\ \citenamefont
  {Parkins}(2008{\natexlab{b}})}]{morrison2008collective}%
  \BibitemOpen
  \bibfield  {author} {\bibinfo {author} {\bibfnamefont {S.}~\bibnamefont
  {Morrison}}\ and\ \bibinfo {author} {\bibfnamefont {A.~S.}\ \bibnamefont
  {Parkins}},\ }\bibfield  {title} {\bibinfo {title} {Collective spin systems
  in dispersive optical cavity qed: Quantum phase transitions and
  entanglement},\ }\href {https://doi.org/10.1103/PhysRevA.77.043810}
  {\bibfield  {journal} {\bibinfo  {journal} {Phys. Rev. A}\ }\textbf {\bibinfo
  {volume} {77}},\ \bibinfo {pages} {043810} (\bibinfo {year}
  {2008}{\natexlab{b}})}\BibitemShut {NoStop}%
\bibitem [{\citenamefont {Cosme}\ \emph {et~al.}(2023)\citenamefont {Cosme},
  \citenamefont {Skulte},\ and\ \citenamefont {Mathey}}]{cosme2023bridging}%
  \BibitemOpen
  \bibfield  {author} {\bibinfo {author} {\bibfnamefont {J.~G.}\ \bibnamefont
  {Cosme}}, \bibinfo {author} {\bibfnamefont {J.}~\bibnamefont {Skulte}},\ and\
  \bibinfo {author} {\bibfnamefont {L.}~\bibnamefont {Mathey}},\ }\bibfield
  {title} {\bibinfo {title} {Bridging closed and dissipative discrete time
  crystals in spin systems with infinite-range interactions},\ }\href
  {https://doi.org/10.1103/PhysRevB.108.024302} {\bibfield  {journal} {\bibinfo
   {journal} {Phys. Rev. B}\ }\textbf {\bibinfo {volume} {108}},\ \bibinfo
  {pages} {024302} (\bibinfo {year} {2023})}\BibitemShut {NoStop}%
\bibitem [{\citenamefont {Finger}\ \emph {et~al.}(2023)\citenamefont {Finger},
  \citenamefont {Rosa-Medina}, \citenamefont {Reiter}, \citenamefont
  {Christodoulou}, \citenamefont {Donner},\ and\ \citenamefont
  {Esslinger}}]{finger2023spin}%
  \BibitemOpen
  \bibfield  {author} {\bibinfo {author} {\bibfnamefont {F.}~\bibnamefont
  {Finger}}, \bibinfo {author} {\bibfnamefont {R.}~\bibnamefont {Rosa-Medina}},
  \bibinfo {author} {\bibfnamefont {N.}~\bibnamefont {Reiter}}, \bibinfo
  {author} {\bibfnamefont {P.}~\bibnamefont {Christodoulou}}, \bibinfo {author}
  {\bibfnamefont {T.}~\bibnamefont {Donner}},\ and\ \bibinfo {author}
  {\bibfnamefont {T.}~\bibnamefont {Esslinger}},\ }\bibfield  {title} {\bibinfo
  {title} {Spin- and momentum-correlated atom pairs mediated by photon exchange
  and seeded by vacuum fluctuations},\ }\href@noop {} {\bibfield  {journal}
  {\bibinfo  {journal} {arXiv}\ }\textbf {\bibinfo {volume} {2303.11326}}
  (\bibinfo {year} {2023})}\BibitemShut {NoStop}%
\end{thebibliography}%
\end{document}